\documentclass[Crown,sageh,times]{sagej}

\usepackage{moreverb,url}

\usepackage[colorlinks,bookmarksopen,bookmarksnumbered,citecolor=red,urlcolor=red]{hyperref}

\usepackage{amsthm,amsmath,amsfonts,amssymb}
\usepackage{graphicx}

\usepackage{mathtools,upgreek,eucal,mathrsfs}
\usepackage{times,makeidx,dcolumn,multirow,amssymb,colortbl,bm,url,color}
\usepackage{epstopdf}
\usepackage{algorithm} 
\usepackage{algorithmic} 
\usepackage{epsfig}

\newtheorem{remark}{Remark}

\newcommand{\bxi}{\bm{\bxi}}

\newcommand{\btheta}{\bm{\theta}}

\newcommand{\bt}{{\bf t}}

\newcommand{\bx}{{\bf x}}
\newcommand{\by}{{\bf y}}
\newcommand{\bY}{{\bf Y}}

\newcommand\BibTeX{{\rmfamily B\kern-.05em \textsc{i\kern-.025em b}\kern-.08em
T\kern-.1667em\lower.7ex\hbox{E}\kern-.125emX}}

\def\volumeyear{2016}
\begin{document}

\runninghead{Christen and Rubio}

\title{Dynamic survival analysis: modelling the hazard function via ordinary differential equations}

\author{J. Andres Christen\affilnum{1} and F. Javier Rubio\affilnum{2}}

\affiliation{\affilnum{1}Department of Statistics,
Centre for Research in Mathematics (CIMAT). Guanajuato, Mexico\\
\affilnum{2}Department of Statistical Science, University College London. London, UK}

\corrauth{F. Javier Rubio
Department of Statistical Science, University College London. London, UK}

\email{f.j.rubio@ucl.ac.uk}

\begin{abstract}
The hazard function represents one of the main quantities of interest in the analysis of survival data. We propose a general approach for parametrically modelling the \emph{dynamics} of the hazard function using systems of autonomous ordinary differential equations (ODEs). This modelling approach can be used to provide qualitative and quantitative analyses of the evolution of the hazard function over time. Our proposal capitalises on the extensive literature of ODEs which, in particular, allow for establishing basic rules or laws on the dynamics of the hazard function via the use of autonomous ODEs. We show how to implement the proposed modelling framework in cases where there is an analytic solution to the system of ODEs or where an ODE solver is required to obtain a numerical solution. We focus on the use of a Bayesian modelling approach, but the proposed methodology can also be coupled with maximum likelihood estimation. A simulation study is presented to illustrate the performance of these models and the interplay of sample size and censoring. Two case studies using real data are presented to illustrate the use of the proposed approach and to highlight the interpretability of the corresponding models. We conclude with a discussion on potential extensions of our work and strategies to include covariates into our framework. Although we focus on examples on Medical Statistics, the proposed framework is applicable in any context where the interest lies on estimating and interpreting the dynamics hazard function.
\end{abstract}


\maketitle

\section{Introduction}\label{sec:intro}

Survival analysis represents a classical area of Statistics which is concerned with the analysis of times to event, potentially subject to censoring. Survival analysis methods have been applied in a number of areas, including medicine, epidemiology, genetics, engineering, and biology, to name but a few. The survival function and the hazard function represent two quantities of interest in this area. The survival function provides information about the probability that an individual or population will survive beyond a certain time point. On the other hand, the hazard function represents the instantaneous failure rate at each time point. From a more practical perspective, the hazard function can be interpreted as a quantification of the risk of the event of interest occurring at a specific time, and the evolution of such risk over time \citep{rinne:2014}.

A number of methods to estimate the survival function have been proposed. From a non-parametric frequentist perspective, the most popular estimators are the Kaplan-Meier estimator \citep{kaplan:1958}, which aims at estimating the survival function, and the Nelson-Aalen estimator \citep{nelson:1972,aalen:1978}, which aims at estimating the cumulative hazard function (which is the integral of the hazard function between $t=0$ and $t=t_0$). From a Bayesian non-parametric perspective, several methods for estimating the survival function based on process priors of different types have been developed (see \citealp{hjort:2010} for an overview of these methods). Parametric methods, using Bayesian and frequentist estimation methods, have also regained popularity in survival analysis thanks to the developments of flexible parametric distributions, as well as spline-based methods (see \citealp{eletti:2022} for a review), and their appealing interpretability. 

Several types of estimators of the hazard function have also been proposed. In the frequentist framework, methods based on kernel density estimation and other data-driven smoothing methods have been proposed to estimate the hazard function in the presence of censoring (see \citealp{rebora:2014} for a review on this literature). 

From a Bayesian nonparametric perspective, several types of estimators of the hazard function based on gamma processes \citep{kalbfleisch:1978}, infinite mixtures \citep{kottas:2006}, piecewise processes \citep{arjas:1994} have been studied (see \citealp{ibrahim:2004} for an overview of Bayesian survival methods in survival analysis). Although flexible, all of these methods lack interpretability, as the shape of the hazard function depends on the number of knots, number of components in the mixture, or smoothing parameters. 
Indeed, in many cases the estimated hazard may exhibit a \textit{wiggly} behaviour \citep{mckeague:2000,hess:1999,kottas:2006}, which complicates the interpretation of such estimates.  

Another type of estimators of the hazard function, which enjoys some popularity in practice, are \textit{piece-wise constant} estimators \citep{friedman:1982}. These estimators impose the condition of constant hazard rates within intervals, and their flexibility depends on the number of intervals and their location. 

Parametric approaches are often criticised as the shape of the estimated hazard function is restricted to a few possibilities, depending on the model. For example, the hazard function associated with the Weibull distribution can only be increasing, decreasing or flat, while the hazard function associated with the lognormal distribution can only be unimodal (up-then-down). Moreover, the value of the Weibull hazard function at time $t=0$ is either $0$ or $\infty$, which is an unrealistic assumption. There exist three-parameter flexible distributions with positive support, such as the power generalised Weibull and generalised gamma, which can capture the basic hazard shapes (increasing, decreasing, unimodal, and bathtub). However, these distributions also impose the condition that the hazard function at time $t=0$ is restricted to be either $0$ or $\infty$ (see \citealp{rubio:2019} for a review on distributions with flexible hazard functions).

Modelling the log-hazard (or log-cumulative-hazard) function using splines has also been considered as an alternative method. However, this requires using a large number of parameters to be able to capture complex shapes, and a careful penalised estimation to obtain smooth curves (see \citealp{eletti:2022} for an overview on these methods). 
More recently, \cite{tang:2022b} proposed a framework for modelling the dynamic change of the cumulative hazard function through an ordinary differential equation, but focused on the inclusion of covariates through this formulation based on known hazard structures. In a more particular setting, \cite{tang:2022a} consider a general ODE form for the cumulative hazard function (allowing for the inclusion of covariates), and estimate it through a feed-forward neural network with the cumulative hazard as an input. Related approaches have been recently explored in \cite{danks:2022}, who used neural
network-based ODEs for modelling cumulative distribution functions. Indeed, these methods tend to focus on adding flexibility to the model, rather than interpreting the hazard or cumulative hazard functions. 


We propose a novel approach for modelling the \emph{dynamics} of the hazard function using systems of first order ODEs, taking the hazard function as a (positive) state variable of the system of ODEs. 
In our context, the word \emph{dynamics} refers to the description of the evolution of the hazard function over time through systems of ODEs.
This modelling approach can also lead to very flexible hazard functions, depending on the system of ODEs employed to characterise the dynamics of the hazard function.
For instance, it is well known that autonomous scalar ODEs have monotonic solutions \citep[][see also Section \ref{sec:examples}]{wallach:1948}. Consequently, using a single ODE to model the hazard function can only capture increasing or decreasing shapes of the hazard function. However, by incorporating additional states, it is possible to obtain non-monotonic hazard shapes, as shown in Section \ref{sec:examples}. These internal states allow us to account for complex interactions, delays, or memory effects \citep{farkas:1984} that would not be possible to model with a scalar ODE (see Section \ref{sec:discussion} for an example). Therefore, the choice of the system of ODEs for modelling the hazard function allows for adding flexibility in an interpretable manner, as the additional hidden states not only enrich the family of hazard function models but they also have a clear dynamic influence on the hazard function itself.
The proposed framework is of particular interest to users aiming at understanding and interpreting the evolution of the hazard function over time, in contrast to scenarios where the only aim is to estimate the survival function at specific time points. That is, the proposed approach is relevant in cases where the user is interested in a quantitative and qualitative analysis of the dynamics of the hazard function.  We capitalise on the vast literature on ODEs, which provides a number of systems of ODEs with interpretable parameters. In particular, we provide examples using autonomous systems of ODEs which facilitate the interpretation of the dynamics of the hazard function and fill a void in the current literature of hazard-based models.
For clarity of exposition and to facilitate a comprehensive analysis of the proposed approach, we will focus on the context without covariates, but we  conclude with a discussion on general strategies for the inclusion of covariates. 

In Section \ref{sec:model}, we present the model formulation and discuss the advantages of modelling the hazard function through autonomous systems of ODEs. 
We present a discussion on modern tools to verify identifiability of hazard models obtained as a solution of systems of ODEs.
We also compare our formulation against previous models for the cumulative hazard function using ODEs \citep{tang:2022a,tang:2022b}. In Section \ref{sec:examples}, we present two particular models for the hazard function using classical systems of autonomous first order ODEs. 
We provide an interpretation of the additional hidden states and the added flexibility obtained through their inclusion. 
Section \ref{sec:inference} presents a discussion on the calculation of the likelihood function associated with the proposed modelling approach, covering the cases where the system of ODEs has an analytic solution or where the use of numerical ODE solvers is required to approximate this solution. Section \ref{sec:simulation} presents a simulation study that illustrates the ability to recover the true values of the parameters and hazard shapes, as well as the effects of sample size and censoring. Section \ref{sec:applications} presents two case studies that illustrate the use of the proposed modelling approach using real data. We conclude with a discussion and possible extensions of this work in Section \ref{sec:discussion}. Software and real data examples can be found at: \url{https://github.com/FJRubio67/ODESurv} for R code, and \url{https://github.com/andreschristen/ODESurv} for Python code.

\section{Modelling the hazard function through ODEs}\label{sec:model}

Let ${\bf o} = \{o_1,\dots,o_n\}$ be a sequence of survival times, $c_i \in \mathbb{R}_+$ be the corresponding right-censoring times, $t_i=\min\{o_i,c_i\}$ be the observed times, and $\delta_i=\mbox{I}(o_i\leq c_i)$ be the indicator that observation $i$ is uncensored, $i=1,\dots,n$.
Suppose that the survival times are generated by an absolutely continuous probability distribution with positive support, and let $f(t)$ denote its probability density function (pdf), $F(t) = \int_0^t f(r) dr$ be the cumulative distribution function (cdf), and $S(t) = 1-F(t)$ be the survival function. From these functions, we can also derive the hazard function $h(t) = -\frac{S'(t)}{S(t)}$, where $S'(t) = \dfrac{d}{dt} S(t)$, and the cumulative hazard function $H(t) = \int_0^t h(r) dr = -\log S(t)$.

We propose parametrically modelling the hazard function $h(\cdot)$ through a system of first order ODEs with initial condition as follows.
Let $q_j: \mathbb{R}^+ \to \mathbb{R}$, $j=1,\dots,m$, be a collection of differentiable functions, and let us denote $\bY(t) = \left(h(t), q_1(t), \ldots, q_m(t) \right)^{\top}$, $t>0$. Define the system of ODEs 
\begin{equation}
\begin{cases}
\bY'(t) =  \psi_{\btheta}(\bY(t), t), \\
 H'(t) = Y_1(t) ,
\end{cases}
\label{eq:general_ode}
\end{equation}
with initial conditions $\bY(0) = \bY_0$ and $H(0) = 0$; taking the initial time $t_0 = 0$ as a simplification, and $Y_1(t) = h(t)$ by definition. 
Let $D \subset \mathbb{R}^{m+2}$ be a closed rectangle.
Assuming that the vector field $\psi_{\btheta}: D \to \mathbb{R}^{m+1}$ is Lipschitz continuous in $\by$ for every $t$ ($\psi_{\btheta}(\by,t)$), and that $\bY_0$ is in the interior of $D$, then there exists a unique solution of the above initial value problem on a vicinity of the initial value (for details see \cite{po-fang1999}, for example). The vector field $\psi_{\btheta}$, its domain $D$, and the initial condition $\bY_0$ are such that the solution for $h(t)$ is positive for all $t>0$, and for any parameter value $\btheta \in \Theta \subset {\mathbb R}^d$. That is, \eqref{eq:general_ode} is a family of systems of ODEs for which one state variable is the hazard function $h$, leading to a family of hazard functions defined by $\btheta$ and $\bY_0$. The cumulative hazard function $H$ is also included in the formulation \eqref{eq:general_ode}, and is thus obtained in the solution of the system. 
The initial condition must satisfy $h(0) \geq 0$, but it is otherwise arbitrary. The initial conditions in \eqref{eq:general_ode} could include $h(0) = h_0 > 0$ indicating that the hazard function takes a non-negative, finite, value at $t=0$, and that all individuals are alive at the start of the follow-up.

While there is no general characterisation of ODE systems guaranteeing positive solutions (specifically, for $Y_1(t)$), several strategies can ensure non-negativity.
One approach involves directly verifying non-negativity through a qualitative analysis of solutions based on the specific structure of the ODE system. This technique is used, for instance, to prove non-negativity in competitive species models, like the Lotka-Volterra equations \citep{boyce:2021}, which we study in Section \ref{sec:examples}. Indeed, a vast literature exists on population dynamics models using ODEs where solutions are naturally positive \citep{hirsch:2012}.

There are also particular theoretical characterisations of systems of ODEs with positive solutions, which are often more restrictive. For example, \cite{bernstein:1999} demonstrates that for autonomous systems (see Section \ref{sec:autonomous}), if the vector field is essentially non-negative (\textit{i.e.}~entry-wise) and the initial conditions are non-negative, then the solutions are non-negative.  
Another strategy involves modelling the dynamics of a transformed hazard function instead of directly modelling the hazard function, such as $\tilde{h}(t) = \log h(t)$, which allows for using vector fields that can be negative. This approach is also used to improve computational efficiency and stability of ODE solvers, as discussed in Sections \ref{sec:examples} and \ref{sec:applications}.

Finally, another option consists of ``forcing'' the solution to become positive \citep{blanes:2022}. This can be done theoretically or numerically by adding a parameter-dependent offset that shifts the solution to the positive quadrant or by discarding parameter values that lead to negative solutions altogether.

In any case, establishing the positivity of state variables in a system of ODEs, as described in \eqref{eq:general_ode}, needs to be done on a case-by-case basis, considering the characteristics of $\psi_{\btheta}$ and the initial conditions $\bY_0$.

Regarding the number of state variables $q_j$'s, or \textit{hidden states},  there is no unique rule for determining the number of $q_j$'s in the ODE system \eqref{eq:general_ode}.
In practice, the number of hidden states to include in \eqref{eq:general_ode} may depend on the biological interpretation of the system. Additional states can also be added to enhance the model's flexibility or incorporate additional effects. In Section \ref{sec:examples}, we present an example of an ODE system with two states which represent the interaction between the hazard function associated to a disease and the response resulting from an intervention (treatment) on the population. In Section \ref{sec:discussion}, we present an example of a system with three states, where the third state is included to model time-delay effects. 

The increased flexibility obtained through additional hidden states is appealing, but as with any parametric model, ensuring parameter identifiability is crucial for making inference about the parameters. Establishing conditions to guarantee identifiability of the solution of systems of ODEs has received a fair amount of attention in the last couple of decades. Studies like \cite{miao:2011} and \cite{qiu:2022} provide conditions to verify identifiability of parameters for families of linear and non-linear systems of ODEs.
Additionally, the Julia package `\texttt{StructuralIdentifiability.jl}' implements state-of-the-art numerical methods to detect non-identifiability. Therefore, we encourage users of our framework to take advantage of these tools to verify identifiability of the parameters of any new system of ODEs. Nonetheless, the literature on systems of ODEs is vast, and the identifiability of many of the classical models (such as those presented in Section \ref{sec:examples}) has already been established.

Before presenting an analysis of this formulation, a remark on the differences of the proposed approach with that proposed in \cite{tang:2022b} and \cite{tang:2022a} seems appropriate. As discussed before, \cite{tang:2022b} proposed modelling the cumulative hazard function via a scalar ODE. That is,
\begin{eqnarray}
\begin{cases}
{H'}(t; \bx) &= \quad \Psi(H(t; \bx ), t ; \bx),\\
H(t_0;\bx ) &= \quad c(\bx),
\end{cases}
\label{eq:chode}
\end{eqnarray}
where $\bx\in{\mathbb R}^p$ are the available covariates. This formulation aims at modelling the dynamics of the cumulative hazard function, which implicitly means modelling the hazard function $h(t; \bx) = {H'}(t; \bx )$, rather than obtaining it as a solution, in contrast to our formulation \eqref{eq:general_ode}. Moreover, our formulation \eqref{eq:general_ode} produces the hazard function and the cumulative hazard function as solutions of a system of ODEs, in contrast to \eqref{eq:chode}. Finally, formulation \eqref{eq:general_ode} allows for including hidden states variables through the inclusion of additional equations in the system. 

On the other hand, note that by setting $h(t \mid \btheta) = \Psi(H(t \mid \btheta ), t )$ and $\Psi(H(t_0 \mid \btheta ), t_0 ) = \bY_0 = h_0$, we obtain an (theoretical) equivalence between our formulation \eqref{eq:general_ode} and formulation \eqref{eq:chode}, restricted to survival models with differentiable hazard functions and models without additional hidden states. More importantly, \cite{tang:2022a} propose to define $\Psi$ through a neural network, with several parameters to estimate. Then, the formulation \eqref{eq:chode} is not seen by \cite{tang:2022a} as an approach to model the (cumulative) hazard function, but simply as a device to use a neural network to construct it.

In sharp contrast, here we focus on \textit{modelling} a family of hazard functions, $h$, through the general formulation in \eqref{eq:general_ode}.  The modelling will take advantage on the knowledge available on the survival problem at hand, capitalising the vast literature on models using systems of ODEs. 

To provide a comprehensive exploration of the proposed modelling approach, we will focus on the case where no covariates are included in the model. Thus, hereafter, we omit the inclusion of covariates in our notation. A comment is included in Section~\ref{sec:discussion} on some strategies for adding covariates to our formulation.

In some cases, the solution to the system of ODEs \eqref{eq:general_ode} is analytic, as we will show in our example in Section \ref{subsec:logistic}. However, this is more the exception than the rule, as many systems of ODEs do not have an analytic solution. In such cases, one can obtain the solution at specific time points $h(t_i)$ by using an appropriate ODE numerical solver, given the vector field $\psi_{\btheta}$ (colloquially known as the ``right hand side'' or rhs), the initial conditions $\bY(0) = \bY_0, H(0) = 0$ and the time points $t_1, \ldots, t_n$ where the solution must be approximated.
Fortunately, the theory of the numerical analysis for systems of ODEs, \textit{i.e.}~initial value problems, is quite robust (see \cite{butcher2016}, for example). Solver implementations include sophisticated implicit step algorithms that dynamically choose methods for stiff and non-stiff systems (\textit{e.g.}~the ``Livermore Solver for Ordinary Differential Equations with Automatic method switching''). Moreover, easy-to-use wrappers of these solvers are available both in R (\texttt{deSolve}, \citealp{soetaert:2010}), and in Python (\texttt{scipy.integrate.solve\_ivp}, \citealp{jones2001}). The examples presented in Section~\ref{sec:examples} are implemented both in R and in Python using these solvers.

\subsection{Autonomous systems of ODEs for modelling the hazard function}\label{sec:autonomous}

If the vector field in the system of ODEs \eqref{eq:general_ode} does not explicitly depend on $t$, it is said to be an autonomous system \citep{boyce:2021}. That is,
\begin{equation}
\begin{cases}
\bY'(t) =  \psi_{\btheta}(\bY(t)), \\
 H'(t) = Y_1(t) .
\end{cases}
\label{eq:aut_ode}
\end{equation}
with $\bY(0) = \bY_0$ and $H(0) = 0$, and $Y_1(t) = h(t)$.
The functions $q_1(t), \ldots , q_m(t)$ are hidden state variables, adding flexibility and richness to the modelling of the hazard function $h$, as shown in the examples in Sections \ref{sec:examples} and \ref{sec:discussion}. 
In fact, \cite{kunze1999} show how a target smooth function may be arbitrarily close to the solution of an autonomous ODE, using an $N$ degree polynomial $\psi(h)$, for $N$ sufficiently large. This result in turn shows that the ODE system \eqref{eq:aut_ode} can be used to produce very flexible hazard functions. 
However, the primary focus of this paper extends beyond simply developing another flexible hazard modeling framework. Instead, it aims to explore the application of autonomous systems of ODEs, as in \eqref{eq:aut_ode}, to \textit{model} the hazard function $h$ in a manner that offers a clear and useful \textit{interpretation} concerning the survival problem under consideration.

As opposed to time dependent non-autonomous ODEs, autonomous ODEs provide qualitative insights regarding the dynamic evolution of the system.
In Section \ref{sec:examples}, we present an example illustrating a qualitative analysis of a system of ODEs, which shows how the solutions evolve over time by examining their equilibrium points.
Since the times of Isaac Newton, autonomous ODEs have been used to model systems by stating principles or laws, namely, the vector field $\psi_{\btheta}$, that remain fixed in time. This powerful tool has been successfully used to model all sorts of phenomena through ``basic rules'', ranging from a variety of areas such as physics, chemistry, biology, epidemics, and sociology, to name but a few \citep{borzi2022}. Our goal is to harness this powerful tool to model hazard functions, for specific problems, based on basic rules dictated by the structure of the system of ODEs. In fact, general qualitative descriptions of hazard functions are already common in survival analysis and reliability \citep{simes:1985,royston:2002,rubio:2019,meeker:2022}. Here we aim to go one step further, turning qualitative assessments into basic interpretable rules, as a tool for modelling hazard functions, as will be explained in the examples in Section \ref{sec:examples}.

\section{Examples of hazard models defined through ODEs}\label{sec:examples}

In this section we present three examples that we consider of practical interest. 
In Section \ref{subsec:1stOODE}, we show that the ODE formulation \eqref{eq:general_ode} can be used to represent common probability distributions with positive support, but also contains extensions, including distributions with a hazard function $h$ with positive and finite initial condition $h(0)$.
A particular model without hidden states, obtained by specifying the hazard function as the solution to the logistic growth model \citep{boyce:2021}, is presented in Section \ref{subsec:logistic}. This model has been largely used in ecology for modelling population growth, and represents a classical example in most textbooks on ODEs. This model does not include hidden state variables $q$ and has an analytic solution.
Section \ref{subsec:HT} introduces a model obtained by defining the hazard function as the solution of a version of the competitive Lotka-Volterra model \citep{kot:2001,murray:2002}. The competitive Lotka-Volterra model is often used to model species in competition, but here we provide an alternative use and interpretation of this model to represent the interaction of the hazard associated with a disease and the response coming from a combination of therapeutic interventions and the immune system (at the population level). This model has a hidden state variable $q$ and no analytic solution, so a numerical solver is employed to approximate its solution. 
This model also illustrates the richness of solutions one can obtain by adding a hidden state to the logistic growth model.

\subsection{Probability distributions defined through a first order ODE}\label{subsec:1stOODE}
Note that any given smooth hazard function ${h(t)}$ may be trivially formulated as the solution of an ODE, with $\psi({h},t) = {h}'(t)$. The following remark shows that our formulation \eqref{eq:general_ode} can be used to represent common absolutely continuous probability distributions with positive support by specifying a Bernoulli-type differential equation \citep{boyce:2021}. 
\begin{remark}
Let $f(t)$ be a differentiable probability density function $f(t)$ with positive support. By differentiating $h(t) = \frac{f(t)}{S(t)}$, it follows that 
\begin{equation}
h'(t) = a(t) h(t) + h(t)^2,
\label{eq:ber}
\end{equation}
where $a(t) = \dfrac{f'(t)}{f(t)} = \frac{d}{dt}\log f(t)$. Equation \eqref{eq:ber} represents a Bernoulli differential equation \citep{baran:1987}. Equation \eqref{eq:ber} can also be seen as a Riccati-type differential equation \citep{onose:1986}.
\end{remark}
The term $a(t)$ in \eqref{eq:ber} may explicitly depend on $t$, which implies that the corresponding ODE is non-autonomous. If the term $a(t)$ in \eqref{eq:ber} can be written in terms of $h(t)$ only, then the corresponding ODE is autonomous. Thus, the term $a(t)$ in \eqref{eq:ber} can be interpreted as an ``autonomy coefficient''.

For example, the Weibull hazard function, $h(t ) = \beta \kappa t^{\kappa -1}$, may be obtained as the solution to the ODE \eqref{eq:ber} with $a(t) =  (\kappa - 1)\left(\dfrac{\beta \kappa}{h(t)}\right)^{\frac{1}{\kappa-1}} - h(t)$, for $\kappa \neq 1$. This is an autonomous ODE. In contrast, for the log-normal distribution $a(t) = \frac{\mu -\sigma^2 -\log (t)}{\sigma ^2 t}$, which implies that the ODE \eqref{eq:ber} associated with the log-normal distribution is non-autonomous.
This is in line with the classical result that, under mild regularity conditions, autonomous scalar ODEs have monotonic solutions \citep{wallach:1948}.
These ideas are summarised in our context in the following remark. 

\begin{remark}\label{th:monotone}
Consider the first order (one-dimensional) autonomous ODE 
\begin{equation*}
h'(t) = \psi(h), \quad h(t_0) = h_0,
\end{equation*}
where $\psi$ is a continuous function, and $h(t)>0$ is the corresponding solution. Then, $h(t)$ is monotonic in $t>0$.
Moreover, Let $h(t)>0$ be the solution to a first order (one-dimensional) ODE 
\begin{equation}
h'(t) = \psi(h,t), \quad h(t_0) = h_0,
\label{eq:FODE}
\end{equation}
If $h(t)$ is monotonic in $t>0$, then the ODE \eqref{eq:FODE} is autonomous. That is, $\psi(h,t)$ depends only on $h$.
\end{remark}
This remark also implies that non-monotonic hazard functions can only be obtained from first order, one-dimensional, non-autonomous ODEs. As we will show later, one can also obtain non-monotonic hazard shapes with \textit{systems} of autonomous ODEs with hidden states.


Some references have claimed that equations of type \eqref{eq:ber} can be seen as a characterisation of absolutely continuous distributions with positive support (see \citealp{baran:1987} and \citealp{onose:1986}). However, our formulation \eqref{eq:general_ode} indicates otherwise, as formulations with hidden state variables cannot be reduced to a Bernoulli ODE of type \eqref{eq:ber}. Moreover, for most of the commonly used distributions, the corresponding hazard function $h(t)$ is the solution to an equation of type \eqref{eq:ber}, with $\lim_{t\to 0} h(t)$ restricted to be either $0$ or $\infty$.
%


\subsection{Logistic growth hazard model}\label{subsec:logistic}

Next, we present an example of \eqref{eq:general_ode} without hidden state variables, and with positive and finite initial conditions on $h$. The logistic growth hazard model is defined through the following system of ODEs,
\begin{eqnarray}
\begin{cases}
h'(t)  =  \lambda h(t) \left(1 - \dfrac{h(t)}{\kappa}\right), & h(0) = h_0\\
H'(t)  =  h(t), & H(0) = 0.
\end{cases}
\label{eq:logisODE}
\end{eqnarray}
where $\lambda > 0$ represents the intrinsic growth rate of the hazard function, $\kappa > 0$ represents the upper or lower bound of the hazard function, and $h_0 > 0$ is the value of the hazard function at $t=0$. The logistic growth ODE is an autonomous Bernoulli ODE of type \eqref{eq:ber}, with $a(t) = \lambda - (1 + \frac{\lambda}{\kappa}) h(t)$.
This ODE has the following analytic solution
\begin{equation*}
h(t \mid \lambda, \kappa, h_0) = \frac{\kappa h_0 e^{\lambda t}}{\kappa + h_0 (e^{\lambda t} - 1)} .
\end{equation*}
The solution $h(t)$, $t \in [0,\infty)$ is positive and finite, and $\lim_{t \rightarrow \infty} h(t) = \kappa$. The corresponding cumulative hazard function is
$$
H(t \mid \lambda, \kappa, h_0) = \dfrac{\kappa}{\lambda }  \log \left(\dfrac{\kappa + h_0 \left(e^{\lambda  t}-1\right)}{\kappa} \right).
$$
It is straightforward to see that its corresponding pdf is
$$
f( t \mid \lambda, \kappa, h_0) =
\frac{\kappa^2 h_0 e^{\lambda t - \kappa/\lambda}}
{\left( \kappa + h_0 (e^{\lambda t} - 1) \right)^2} .
$$

\begin{figure}
\begin{center}
\includegraphics[scale=0.5]{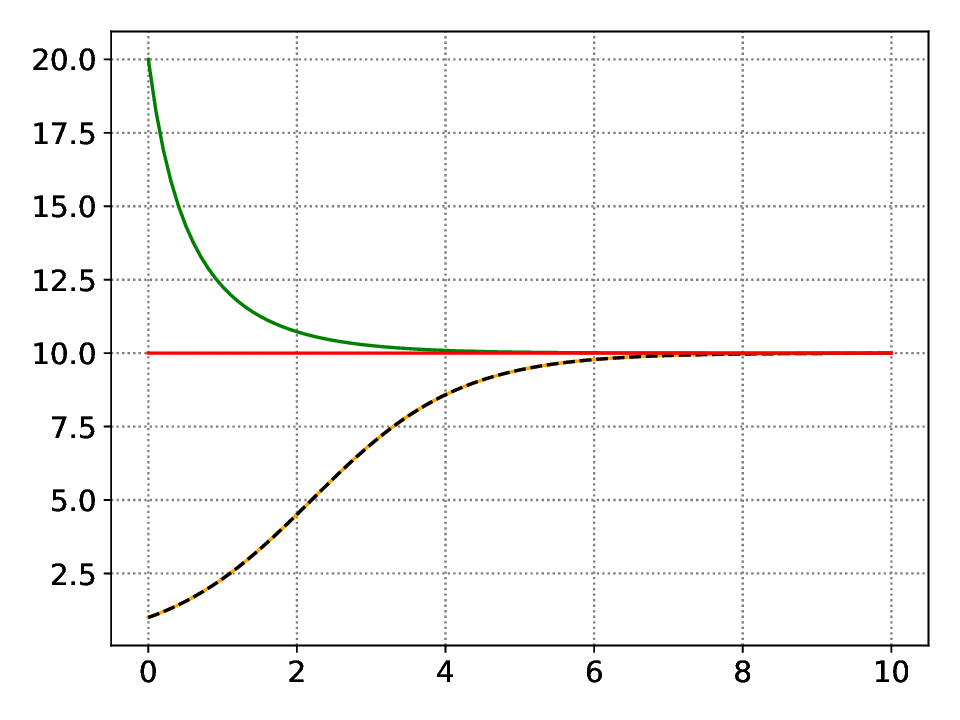}
\caption{Three examples of the logistic growth hazard function, $\lambda = 1, \kappa= 10$. Increasing, $h_0 = \kappa/10$ (orange), decreasing, $h_0 = 2 \kappa$ (green), and constant, $h_0 = \kappa$ (red). The analytic solution is plotted for $h_0 = \kappa/10$ (dashed black line), for comparison.}
\label{fig:logisticHazard}
\end{center}
\end{figure}

Simulating from this model can be done by inverting the cdf, as follows. Let $u$ be realisation of a $U(0,1)$ distribution. Then, 
\begin{equation*}
t = \dfrac{1}{\lambda}\log \left[ 1 + \dfrac{\kappa}{h_0}\left\{ \exp\left( - \dfrac{\lambda}{\kappa}\log(1-u) \right) -1 \right\}\right],
\end{equation*}
is a realisation of the logistic growth model.
An advantage of model \eqref{eq:logisODE} is that it does not force the hazard function to be either $0$ or $\infty$ at $t=0$, in contrast to the Weibull distribution (among other distributions). Moreover, the shapes of the hazard function are also increasing or decreasing (see Figure \ref{fig:logisticHazard}). The parameters have a clear interpretation as $h_0$ represents the value of the hazard function at $t=0$, $\kappa$ represents the asymptotic point of stability of the hazard function, and $\lambda$ represents the growth rate of the hazard function. Figure \ref{fig:logisticHazard} shows some examples of the logistic growth hazard model. The numerical solutions are only presented for illustration, since there exists an analytic solution. The numerical solutions were calculated using the ``Livermore Solver for Ordinary Differential Equations with Automatic method switching'' (\texttt{LSODA}, SciPy implementation). The analytic solution is plotted for $h_0 = \kappa/10$, for comparison.

In all cases, we can notice how $h(t) \rightarrow \kappa$ as $t\to\infty$.

\subsection{Hazard-Response model}\label{subsec:HT}
The hazard-response model is defined through the system of ODEs:
\begin{eqnarray}
\begin{cases}
h'(t)  =  \lambda h(t) \left(1 - \dfrac{h(t)}{\kappa}\right) - \alpha q(t) h(t), & h(0) = h_0 \\
q'(t) =  \beta q(t) \left( 1- \dfrac{q(t)}{\kappa} \right) -\alpha q(t) h(t)  ,  &  q(0) = q_0 \\ 
H'(t)  =  h(t), & H(0) = 0,
\end{cases}
\label{eq:hazardresponse}
\end{eqnarray}
with $\lambda>0$, $\alpha \geq 0$, $\beta>0$, $\kappa>0$, $h_0>0$, and $q_0>0$.
This model represents a particular version of the competitive Lotka-Volterra model \citep{kot:2001,murray:2002}. 
This model formulation assumes that the hazard function $h(t)$ is in competition with a response $q(t)$ (associated with the immune system and any treatment or intervention at the population level).  
In the absence of competition ($\alpha=0$), $h(t)$ follows a logistic-growth and reaches its carrying capacity $\kappa$ as $t\to \infty$. An analogous argument applies on $q(t)$ for the case when $\alpha=0$. When $\alpha>0$, the competition between these two components is modeled through the term $\alpha q(t) h(t)$, and both terms are affected negatively (as the sign is negative in both equations).
The term $q(t)$ may be regarded as a hidden, unobserved variable. In that sense, its units are arbitrary and, without loss of generality, we may fix its carrying capacity equal to the carrying capacity of $h(t)$, namely $\kappa$. The competition term could include a different coefficient for each equation but, as a parsimonious modelling approach, we use the same coefficient ($\alpha$) in both equations. The fully parametrised model is studied in \cite{murray:2002} in the context of ecological modelling as a generalisation of the Lotka-Volterra predator-prey model.

The qualitative analysis of this system is more complicated than that of the simpler logistic-growth model studied in Section~\ref{subsec:logistic}, and may be found in \cite{murray:2002}, Section 3.5. Briefly, this model has four steady state solutions. These steady states can be found by solving the equation $\psi_{\btheta}(h, q) = \mathbf{0}$, which defines four cases: Case 1, $h=0, q=0$; Cases 2 and 3, $h=\kappa, q=0$ and $h=0, q=\kappa$, respectively; and Case 4, $h=h^*, q=q^*$, if both positive, where
$$
h^* = \kappa \left( \frac{1-\alpha \kappa \lambda^{-1}}{D} \right)
\quad \text{and} \quad
q^* = \kappa \left( \frac{1-\alpha \kappa \beta^{-1}}{D}  \right),
$$
with $D=1 - \frac{(\alpha \kappa)^2}{\lambda \beta}$. Case 4 is only relevant when $D \neq 0$. The next step is to investigate if these cases are attractors, that is, if $h(t), q(t)$ tend to one of these points as $t\to\infty$, or not, and under what conditions \citep[see][Section 3.5 for details]{murray:2002}.

Case 1 is not possible as we are assuming $h_0 >0$ and $q_0 >0$. 
It is not possible that $h^*<0$ and $q^*<0$ simultaneously.
If $\alpha \kappa \lambda^{-1} <1$ and $\alpha \kappa \beta^{-1} >1$, then Case 2 is an attractor, that is, the hazard $h(t)$ reaches its carrying capacity as $t\to\infty$ and the response $q(t)$ ``losses'' the competition ($q(t) \to 0$).
Conversely, if $\alpha \kappa \lambda^{-1} > 1$ and $\alpha \kappa \beta^{-1} < 1$, then Case 3 is an attractor, that is, the response $q(t)$ ``wins'' the competition ($q(t) \to \kappa$) and the hazard ``losses'' the competition ($h(t) \to 0$). Case 4 is specially interesting; since both $h^* >0$ and $q^* > 0$, the hazard remains positive ($h(t)\to h^* >0$), although not at its maximum (or minimum), but in an equilibrium with the response ($h^* < \kappa$). However, it is only an attractor when $D>0$. We will say the hazard-response model is in equilibrium if it is in Case 4 with $D>0$.
If $D<0$ then Case 2 or Case 3 are the attractors, depending on the initial conditions.
We will come back to this classification when analysing the real data using this hazard function in Section \ref{sec:applications}.

Given that model \eqref{eq:hazardresponse} does not admit an analytic solution, it is not possible to invert the corresponding cdf directly to simulate samples from this model. Alternatively, we propose an algorithm to approximately simulate from this model using the output from an ODE solver. Algorithm \ref{alg:simHT} shows the steps in the approximate simulation process. The quality of the approximated simulated samples depends on the number of points $M$, and the choice of an appropriate maximum value $t_M$.

\begin{algorithm}[ht]
\caption{Approximate Simulation from the Hazard-Response model \eqref{eq:hazardresponse}}
For given values of the parameters and for a sequence of $M$ time points $\tilde{\bt} = \{t_{1},\dots,t_{M}\}$:
\begin{itemize}
\item[1.] Obtain a numerical solution of the ODE system \eqref{eq:hazardresponse} at $\tilde{\bt}$, using an ODE Solver.
\item[2.]  Fit a monotonic spline basis with coordinates given by the grid $\tilde{\bt}$ and the solutions for the cumulative hazard $H$ at $\tilde{\bt}$. This produces an approximation, $\tilde{H}^{-1}$, of $H^{-1}$.
\item[3.] Generate $u\sim U(0,1)$.
\item[4.] Calculate $t^* = \tilde{H}^{-1}\left( -\log(u) \right)$.
\end{itemize} 
\label{alg:simHT}
\end{algorithm}

\section{Inference}\label{sec:inference}

Since the proposed hazard models are parametric, the log-likelihood function can be written in terms of the hazard function and the cumulative hazard function, as usual:
\begin{equation*}
\ell(\btheta, \bY_0) =  \sum_{i=1}^n   \delta_i \log h(t_i \mid \btheta, \bY_0) - \sum_{i=1}^n H(t_i \mid \btheta, \bY_0),
\end{equation*}
where we are including the initial conditions $\bY_0$ as unknown parameters in this equation. In some cases, as we will discuss in the simulation study and the applications, in Sections \ref{sec:simulation}-\ref{sec:applications}, it is possible to accurately estimate the initial conditions. However, in other cases, as discussed in Section \ref{sec:application2}, the data contains little information about these parameters, leading to practical non-identifiability, which is reflected as flat likelihood surfaces \citep{cole:2020}. Indeed, the initial conditions are commonly fixed at reasonable values in many practical applications of ODE models \citep{simpson:2022}.
Once a solution (analytic or numerical) to the system of ODEs \eqref{eq:general_ode} has been obtained, we can evaluate this likelihood function at specific values of the parameters $\btheta$.
In particular, for systems of ODEs without analytic solution and once a solver has been specified, we can retrieve the hazard and cumulative hazard functions, $h(t\mid \btheta, \bY_0)$ and $H(t\mid \btheta, \bY_0)$, at the sequence of observed time points $\bt = \{t_1,\dots,t_n\}$.
This allows for calculating the maximum likelihood estimates (MLEs) of the parameters $\btheta$ and the initial conditions $\bY_0$ using general-purpose optimisation algorithms (\textit{e.g.}~\texttt{optim} or \texttt{nlminb} in R).

Although it is not possible to come up with general prior choices for the parameters of the hazard models obtained via \eqref{eq:general_ode}, as different models contain different types of parameters, the ease of interpretation of those parameters facilitates choosing either informative or weakly informative priors. The use of improper priors would require a case-by-case analysis to check the propriety of the posterior distribution.

Since the likelihood function can be evaluated at each parameter value, either for analytic or numerical solutions, any general-purpose sampler can be coupled with the proposed models. These include general MCMC samplers such as Metropolis-within-Gibbs sampler (\texttt{BUGS}, \texttt{spBayes}), Hamiltonian Monte Carlo (\texttt{Stan}), or other \textit{ad-hoc} samplers (\texttt{twalk}, \texttt{MCMCPack}).

\section{Simulation study}\label{sec:simulation}

This section presents a simulation study that aims at illustrating the ability to recover the true values of the parameters and hazard shapes, as well as the interplay of sample size and censoring on such aim. We present two simulation scenarios using models \eqref{eq:logisODE} and \eqref{eq:hazardresponse}, with a combination of sample sizes and censoring rates, as described below. 

\subsection{Simulation scenarios}
For simulation scenario 1, we simulate $M=250$ samples of sizes $n=250,500,1000,5000$ from the logistic growth hazard model \eqref{eq:logisODE} with parameters $(\lambda,\kappa,h_0) = (0.5,0.05,3.5)$. These parameter values produce a decreasing hazard function, which starts at $h_0=3.5$, and decreases at a relatively fast rate, $\lambda = 0.5$, to the carrying capacity value $\kappa = 0.05$. These samples are simulated using the procedure described in Section \ref{subsec:logistic}. 
Administrative censoring times (\textit{i.e.}~fixed) are used to induce censoring rates of approximately $50\%$ and $25\%$.    

For simulation scenario 2, we simulate $M=200$ samples of sizes $n=250,500,1000,5000$ from the hazard-response model \eqref{eq:hazardresponse} with parameters $(\lambda,\kappa,\alpha,\beta) = (1.8,0.1,6,4.8)$, using Algorithm \ref{alg:simHT}. The grid $\tilde{\bt}$ is defined with equally spaced values on the interval $[0,150]$, with step size $0.001$. In step 2 of Algorithm \ref{alg:simHT}, we use monotone cubic interpolation. 
We remind the reader that this algorithm produces approximate samples from model \eqref{eq:hazardresponse}. Thus, the aim of this scenario is to assess the ability to recover the true values of the parameters and the shape of the hazard using this simulation strategy. Administrative censoring times are used to induce censoring rates of approximately $50\%$ and $25\%$. 
Rather than estimating the initial conditions $h_0$ and $q_0$, we fix them at specific values $q_0 = 10^{-2}$ and $q_0=10^{-6}$ (in this case, and for illustration purposes, at the true parameter values). These values are inspired by the estimates obtained in case study II, Section \ref{sec:application2}. These parameter values produce a unimodal hazard that stabilises after the mode (stability Case 4). In case study II, Section \ref{sec:applications}, we will discuss how to use prior information to choose the initial conditions at reasonable values. 

For all models we choose weakly informative priors, as follows. For all positive parameters, we adopt gamma priors with scale parameter $2$ and shape parameter $2$. This prior has mean $4$, variance $8$, it accumulates $95\%$ of the probability in the interval $(0.5,11.2)$, and it vanishes at $0$ (which helps repelling the MCMC samplers from visiting regions near zero, that may cause numerical problems). 

For scenario 1, and for each simulated sample, we obtain a posterior sample of size $1,000$ using the t-walk sampler \citep{christen:2010} in R, with a burn-in period of $5,000$ iterations and a thinning period of $50$ iterations (\textit{i.e.}~$55,000$ iterations in total). For scenario 2, and for each simulated sample, we obtain a posterior sample of size $1,000$ using the adaptive Metropolis-within-Gibbs sampler implemented in the R package \texttt{spBayes}, with a burn-in period of $5,000$ iterations and a thinning period of $50$ iterations, and with approximately $0.44$ marginal acceptance rates. In order to obtain a numerically stable solution of the system of ODEs, we first reformulate the system in the logarithmic scale (that is, in terms of $\log h(t)$, see the Appendix) and then apply the Livermore Solver for Ordinary Differential Equations (LSODE) solver from the R package \texttt{DeSolve} \citep{soetaert:2010}. The Jacobian used in the implementation of LSODE is calculated explicitly (see the Appendix) for a better performance of the ODE solver. 

\subsection{Results}
\begin{figure}[ht]
\centering
\begin{tabular}{c c c c}
\multicolumn{4}{c}{$25\%$ censoring}\\
    \includegraphics[width = 3.5cm, height = 3cm]{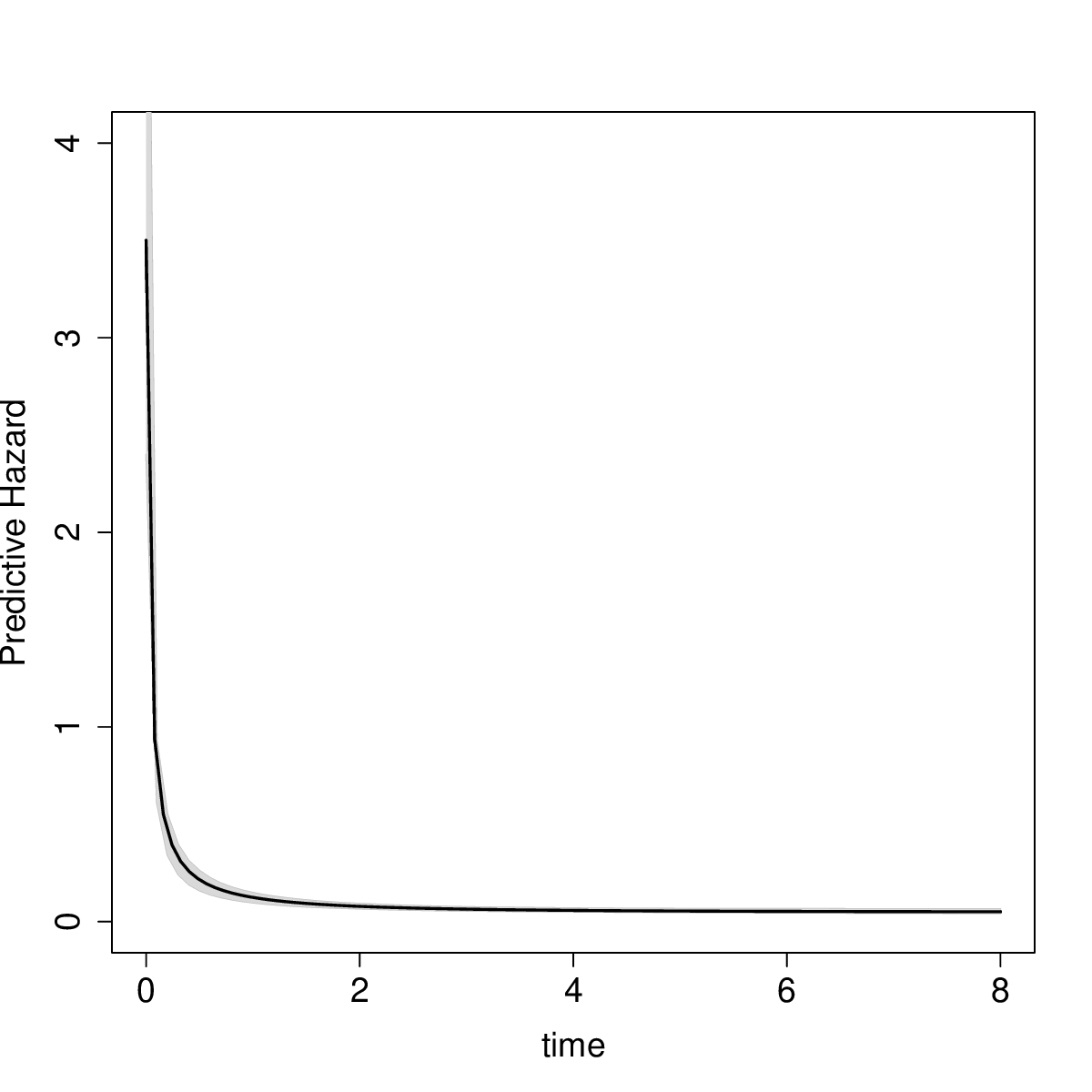} &
    \includegraphics[width = 3.5cm, height = 3cm]{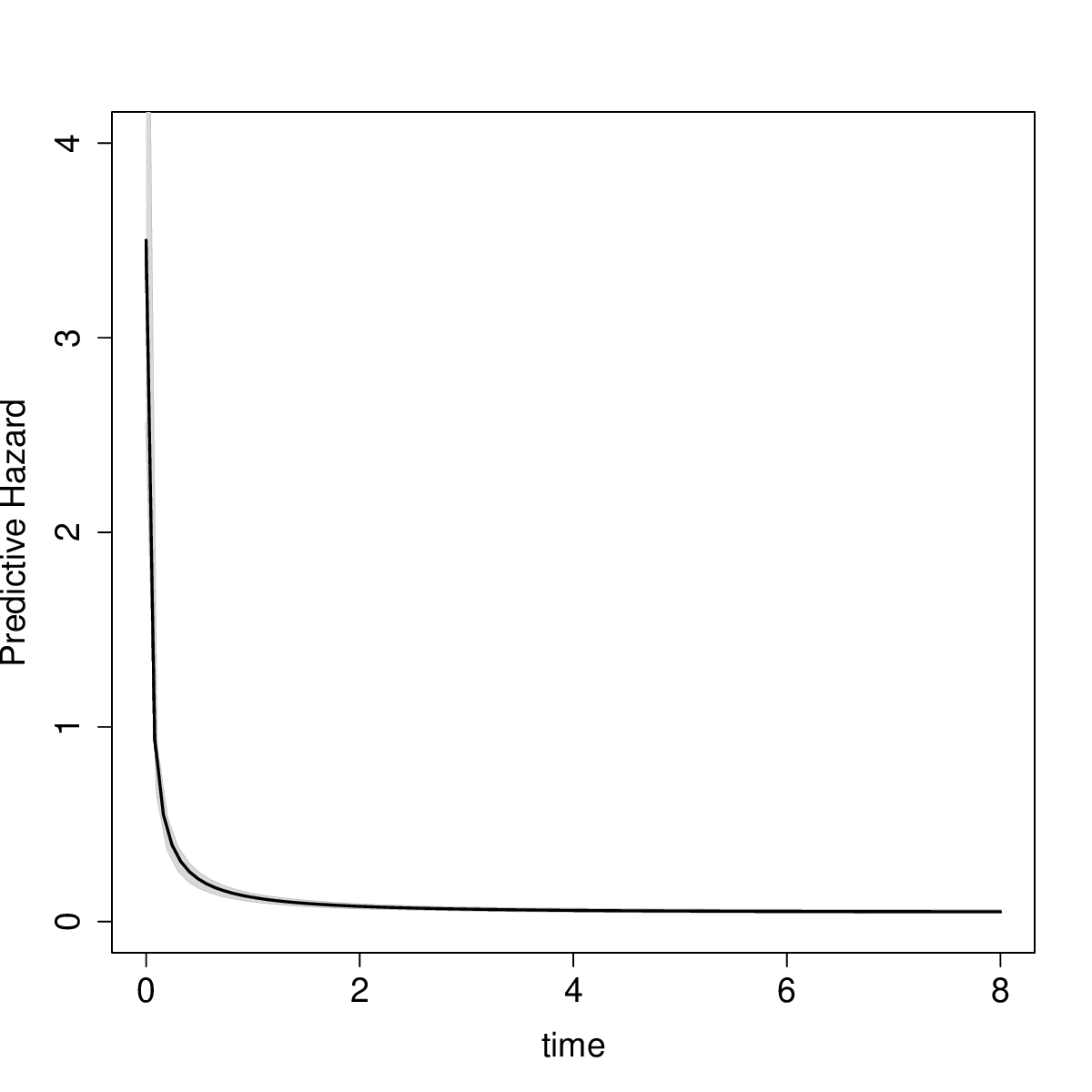} &
    \includegraphics[width = 3.5cm, height = 3cm]{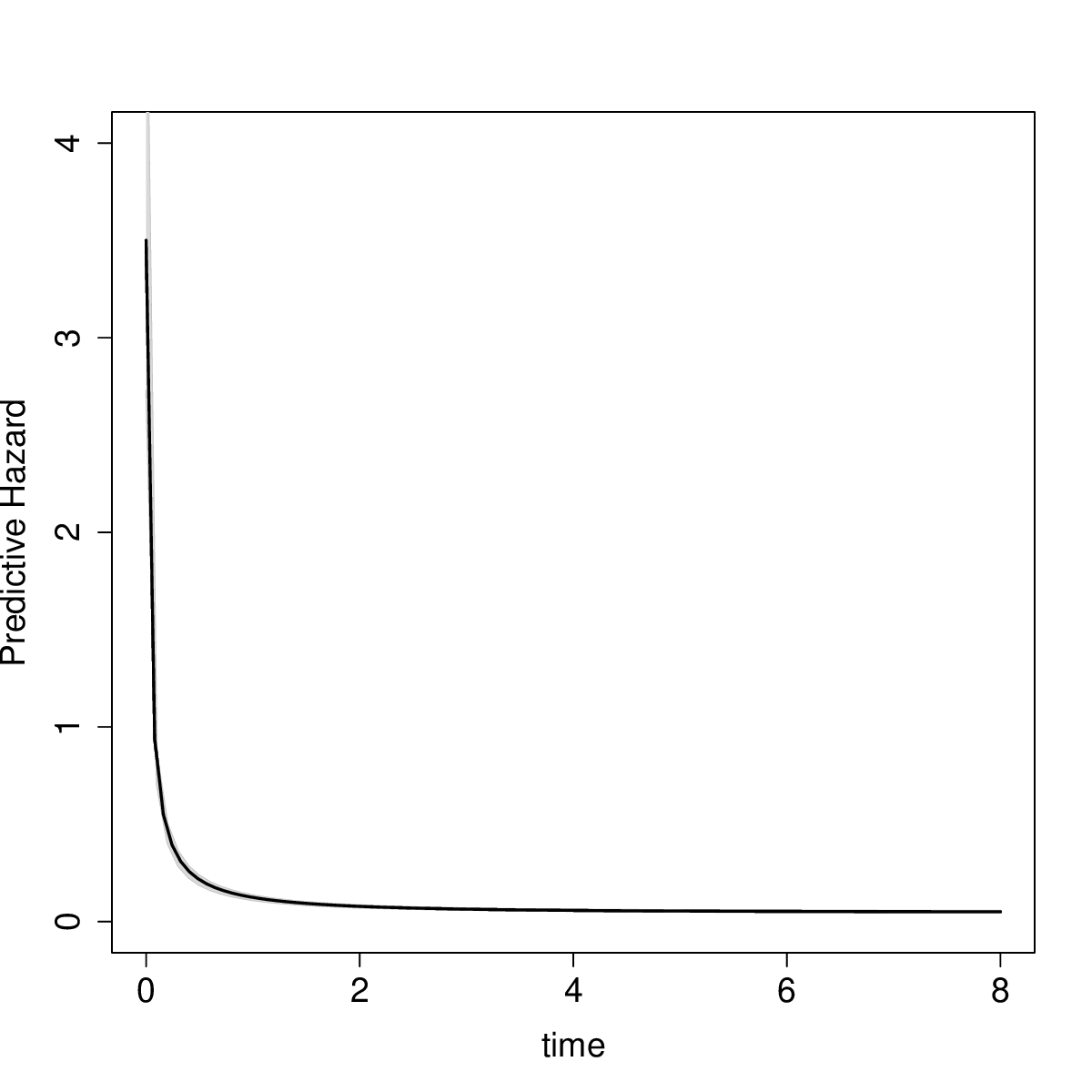} &
    \includegraphics[width = 3.5cm, height = 3cm]{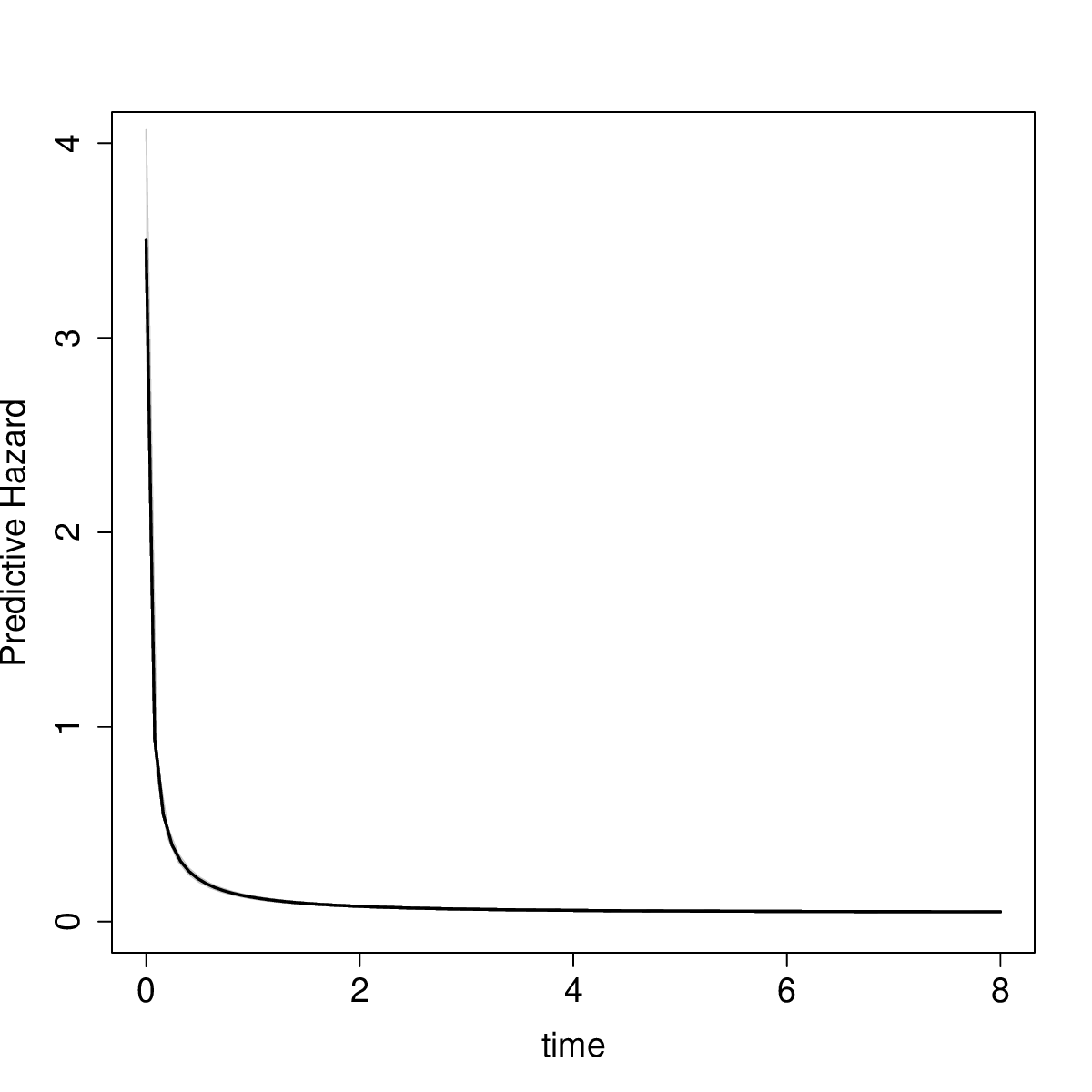} \\
    (a) $n=250$ & (b) $n=500$ & (c) $n=1000$ & (d) $n=5000$ \\
    \multicolumn{4}{c}{$50\%$ censoring}\\
    \includegraphics[width = 3.5cm, height = 3cm]{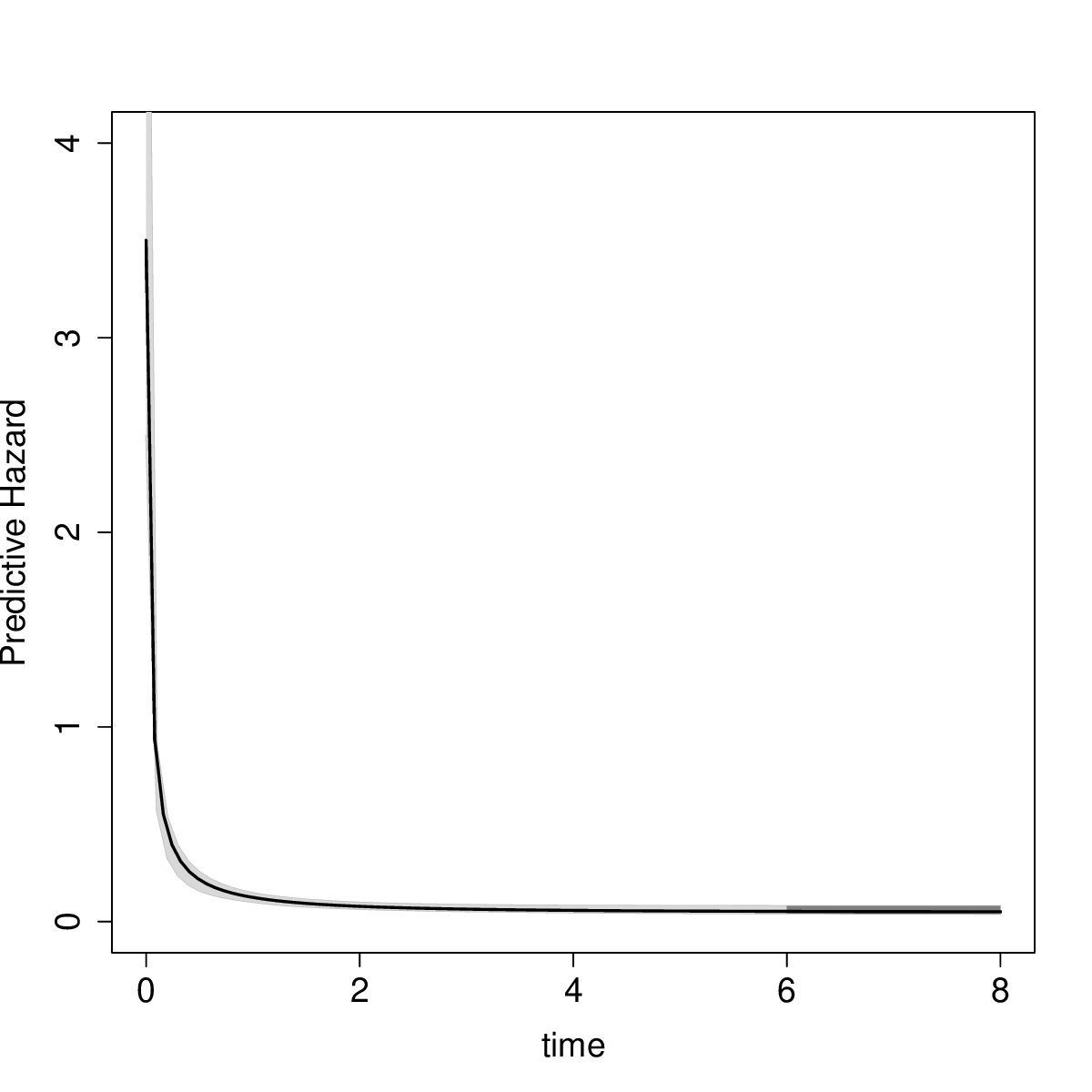} &
    \includegraphics[width = 3.5cm, height = 3cm]{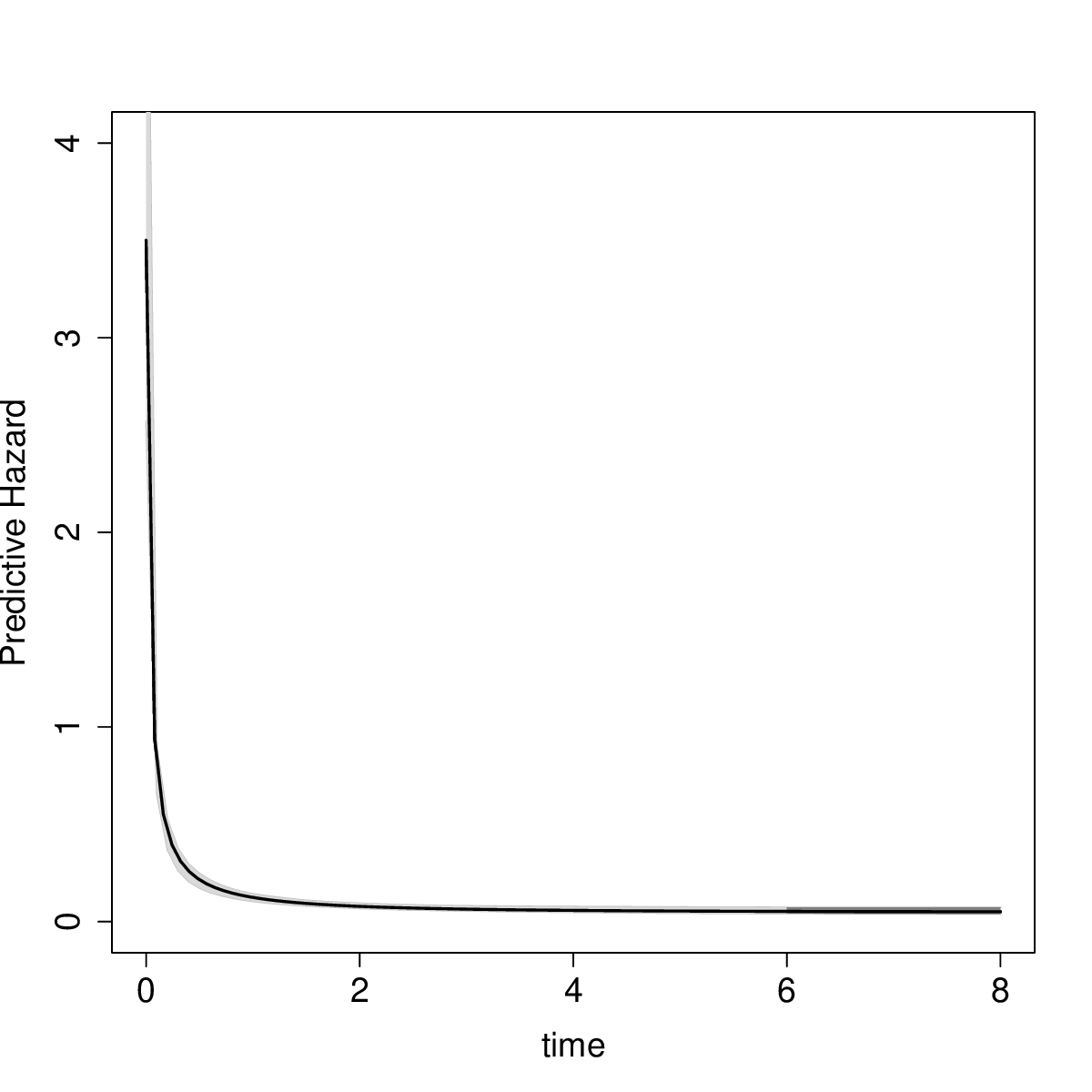} &
    \includegraphics[width = 3.5cm, height = 3cm]{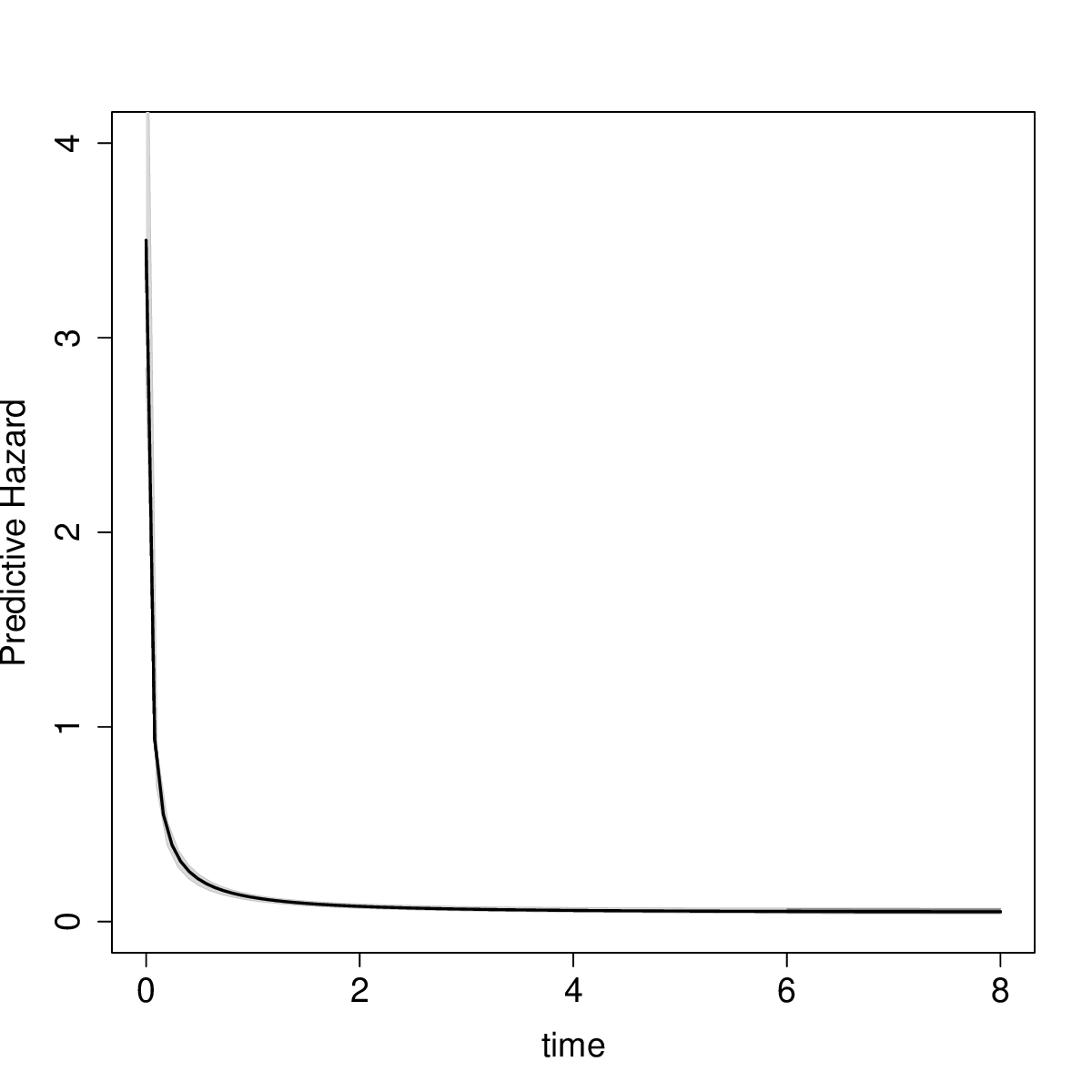} &
    \includegraphics[width = 3.5cm, height = 3cm]{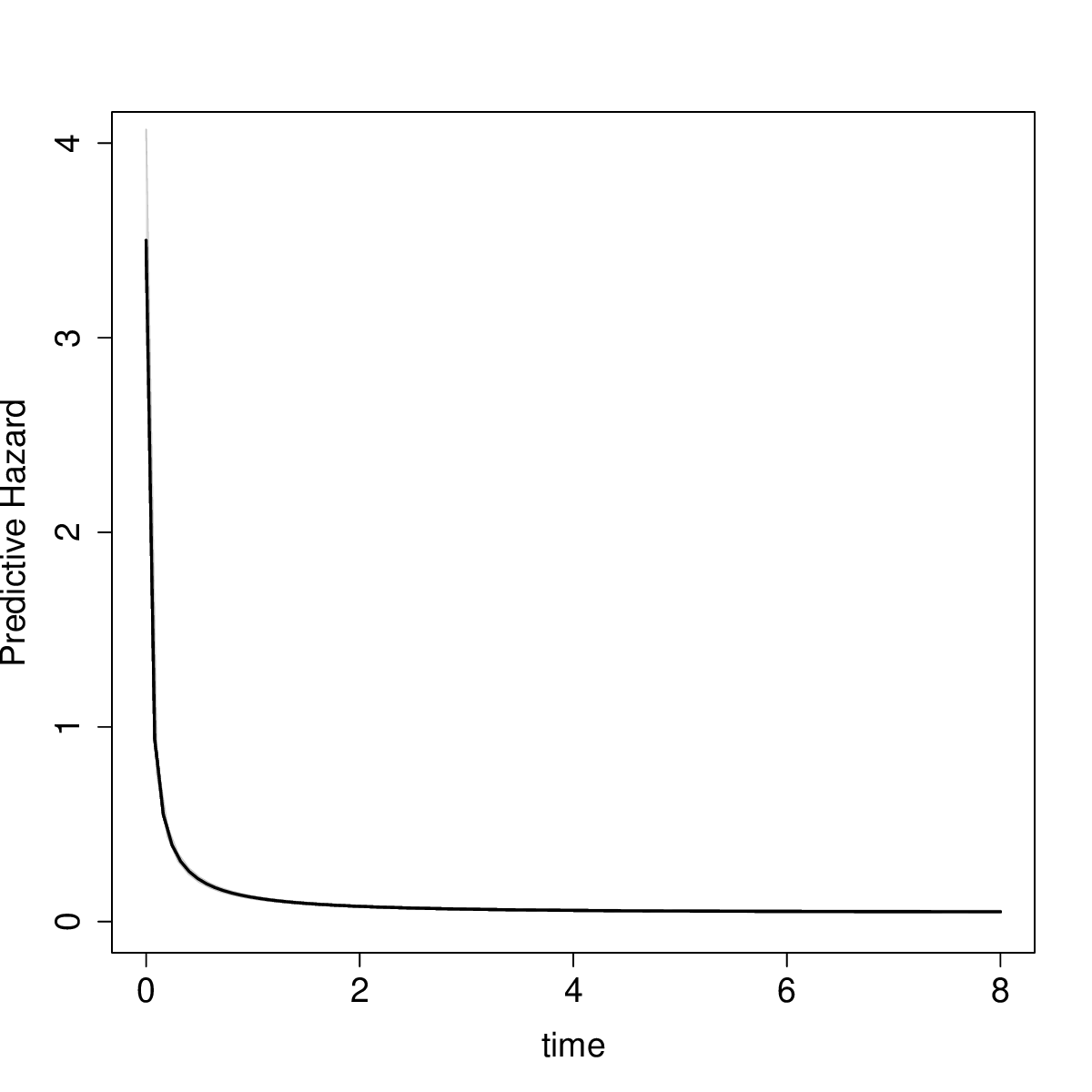} \\
    (e) $n=250$ & (f) $n=500$& (g) $n=1000$ & (h) $n=5000$
    \end{tabular}
    \caption{Simulation scenario 1: Predictive hazards and $95\%$ predictive intervals. The dark gray area indicates the censoring point.}
    \label{fig:predhazS1}
\end{figure}

Tables \ref{tab:S1C25}--\ref{tab:S1C50}
in the Appendix show summaries of the results obtained for simulation scenario 1. These tables show the average posterior mean (across the simulated samples), average posterior median, average posterior standard deviation, average RMSE with respect to the posterior mean, and coverage of the $95\%$ credible intervals. We can see that it is possible to recover all of the parameters, in the sense that the empirical bias of point Bayes estimates (posterior mean and posterior median) is reduced as the sample size increases or the censoring rate decreases. The empirical coverage is close to the nominal value for all parameters. From these tables, we can also see that, unsurprisingly, the estimation of the initial condition $h_0$ is the most challenging (see the RMSE and SD values). Figure \ref{fig:predhazS1} shows the true hazard function together with the $95\%$ predictive intervals, which are narrow, even for the smallest sample size considered (and while adopting generic weakly informative priors). 

Tables \ref{tab:S2C25}--\ref{tab:S2C50}
show the summaries of the results for simulation scenario 2. We can see that the Bayes estimates are close to the true values of the parameters, but exhibit some degree of bias which does not disappear with increasing sample size. This seems to be a result of the step size used in the simulation and the interval length. In order to reduce this bias, it is required to use a smaller step size and larger interval length. On the other hand, we can see that this configuration allows one to recover the shape of the hazard with high accuracy, as shown in Figure \ref{fig:predhazS3} in the Appendix, which presents the true hazard function together with the $95\%$ predictive intervals.

For benchmarking purposes, we compare the predictive hazard functions obtained in simulation scenario 2 and a B-spline estimator of the hazard function (using the R package \texttt{bshazard}) against the true hazard function. For this purpose, we use the restricted $L_1$ distance between two hazard functions, $h$ and $\tilde{h}$, proposed in \cite{de:2009}:
\begin{equation*}
d_R(h_0,\tilde{h}) = \int_0^{t^*} \vert h_0(t) - \tilde{h}(t) \vert dt.
\end{equation*}
We take $t^*$ to be the maximum follow up time for each case. 
We calculate this distance in two cases: when $\tilde{h}$ represents the predictive hazard function obtained in simulation scenario 2, and when $\tilde{h}$ is the B-spline estimator of the hazard function, with $h_0$ being the true hazard function. The corresponding hazard functions are approximated by interpolating the solutions retrieved by the ODE solver and the R package \texttt{bshazard} on a grid of 200 equidistant points on $[0, t^*]$ using cubic splines. These approximations are then integrated using the R command \texttt{integrate}.
Figures \ref{fig:dist25}-\ref{fig:dist50}
in the Appendix display the violin plots corresponding to the $250$ distances obtained in each case. It is evident that the variability of the B-spline estimator consistently results in higher distances from the true generating model.

\section{Real data applications}\label{sec:applications}
In this section, we present two case studies using real data on Leukemia and Breast cancer. The first case study concerns the analysis of the \texttt{LeukSurv} data set from the R package \texttt{spBayesSurv} using the logistic ODE hazard model \eqref{eq:logisODE}. This represents an example using a hazard model with analytic solution, and where one can estimate all of the model parameters, including the initial conditions. The second case study concerns the analysis of the \texttt{rotterdam} breast cancer data set from the R package \texttt{survival} using the hazard-response model \eqref{eq:hazardresponse}. This represents a case where no analytic solution exists, and thus we need to employ numerical ODE solvers to approximate the solution. Moreover, due to the evolution of breast cancer (the first time-to-event is observed around 6 weeks after the start of follow-up), the data contain no information regarding the initial conditions. We use the interpretation of the model parameters and previous information about this type of cancer in the literature to choose these values. 

\subsection{Case study I: Leukemia data}\label{sec:application1}
In this application, we analyse the \texttt{LeukSurv} data set from the \texttt{spBayesSurv} R package, using the logistic hazard model \eqref{eq:logisODE} presented in Section \ref{subsec:logistic}. This data set contains information about the survival of $n=1043$ patients with acute myeloid leukemia. The maximum follow-up time was $13.6$ years for this data set. By the end of the follow-up $879$ individuals died and $164$ were still alive.

\begin{figure}
\begin{center}
\begin{tabular}{c c}
\includegraphics[scale=0.35]{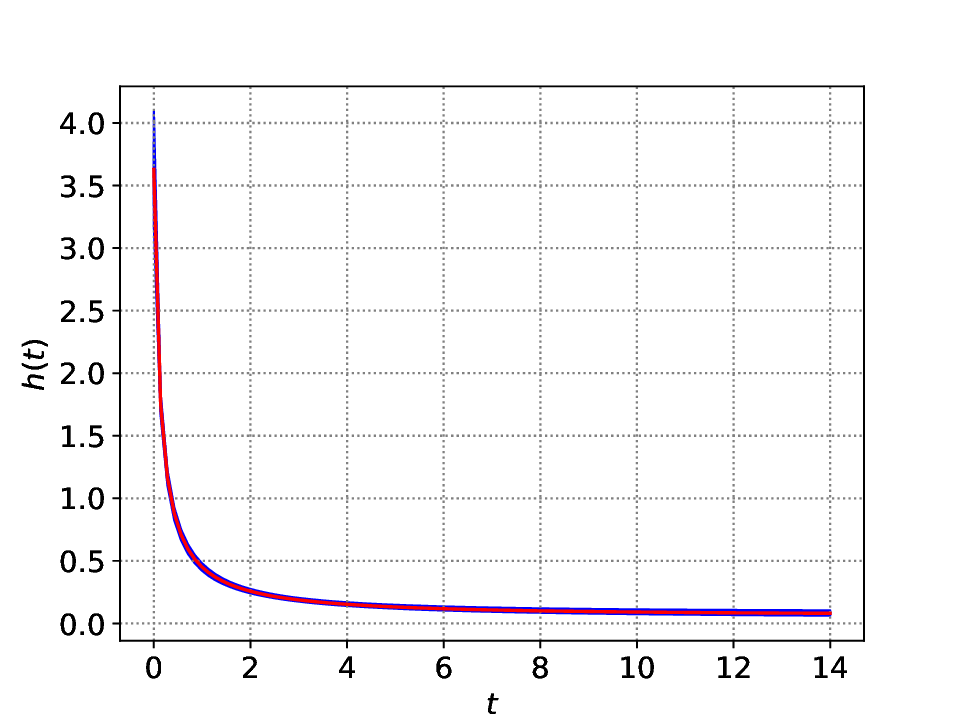} &
\includegraphics[scale=0.35]{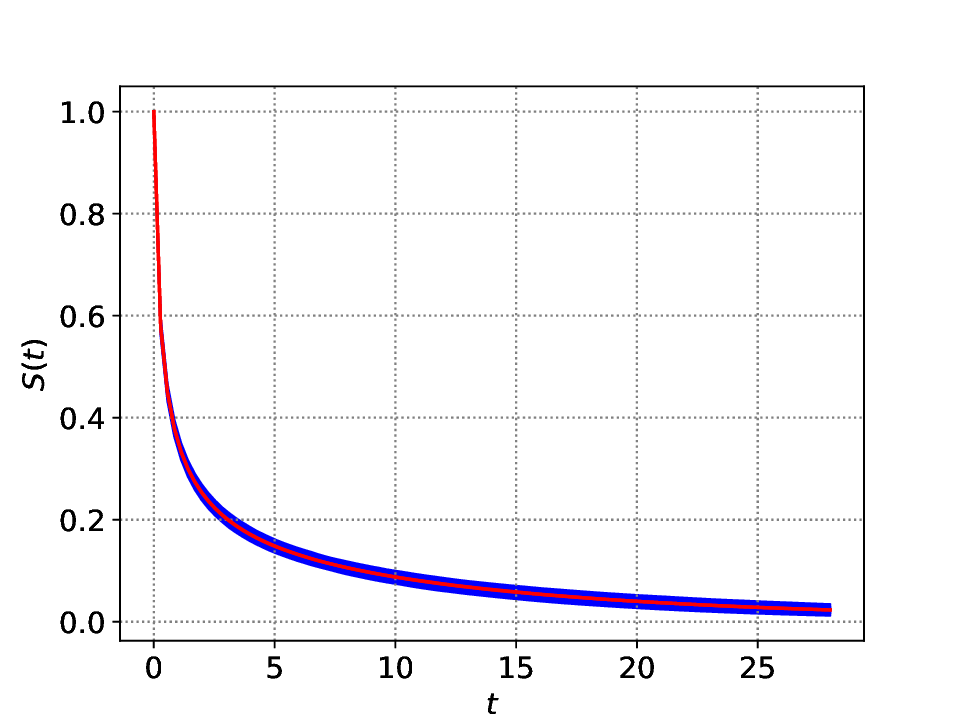} \\
(a) & (b)
\end{tabular}
\end{center}
\caption{\texttt{LeukSurv} data: (a) posterior predictive hazard function, and (b) survival functions}
 \label{fig:logsitic_h_S}
\end{figure}

We simulate from the posterior distribution of the parameters $(\lambda,\kappa,h_0)$ of the logistic hazard model \eqref{eq:logisODE} using the t-walk sampler \citep{christen:2010}. The t-walk is a self-adaptive MCMC algorithm that requires no tuning parameters. 
We obtain $500,000$ posterior samples, and use a burn-in period of $20,000$ iterations and a thinning period of $167$ iterations. This is the estimated integrated auto-correlation time \citep[IAT, see details of output analysis in][]{Molina2022}, and divided by the dimension (here, three) results in $55.6$, leading to an effective sample size of $480,000/167 \approx 2870$. 
Figure \ref{fig:logsitic_h_S} shows the corresponding posterior predictive hazard and survival functions for this population of cancer patients. The posterior predictive hazard function starts at $h_0$ (posterior mean $\approx 3.6$) and sharply decreases to $\kappa$ ($95\%$ posterior credible interval $(0.012,0.125)$) within the first 4 years. These results are in line with the progression of this type of cancer tumours, which tend to develop quickly with only a small portion of the population responding positively to treatments \citep{shallis:2019}, leading to poor prognosis and survival. 
Now, comparing the logistic hazard model against the Weibull distribution using the Bayesian information criterion (BIC): 
\begin{equation*}
BIC = k \log(n) - 2 \ell(\lambda,\kappa,h_0) ,  
\end{equation*}
where $k=3$, the total number of unknown parameters, and $n$ is the sample size.
the logistic hazard model has a $BIC = 1862.6$, while fitting a Weibull distribution to the same data we obtain $BIC = 1898.5$. Thus, the data clearly favours the logistic hazard model. 

\subsection{Case study II: Breast cancer data}\label{sec:application2}
In this application, we analyse the \texttt{rotterdam} data set \citep{royston:2013} from the \texttt{survival} R package. This data set contains information about the survival of $n=2982$ breast cancer patients, from which $1272$ died within the maximum follow-up period ($19.3$ years). It is known that some of these patients received hormonal treatment, chemotherapy, and/or surgical treatment. 
We expect that the evolution of the population hazard function over time to depend on the response to the treatments and the natural immunological response. Thus, we model the survival times using the hazard-response model \eqref{eq:hazardresponse}. The minimum survival time in this data set is $45$ days, thus, we do not expect to have information about the initial conditions $h_0$ and $q_0$. We have verified this assumption by fitting the model where the initial conditions are assumed to be unknown parameters, and we found that the posteriors are virtually the same as the priors, indicating weak identifiability \citep{cole:2020} of $h_0$ and $q_0$. Consequently, we fix these values using prior knowledge. Indeed, fixing the initial conditions in ODE models to some reasonable values is a common practice \citep{simpson:2022}. It is known that the prognosis of breast cancer is relatively good compared to other cancers, and that the mortality rate during the first few months is very low. Based on this prior information, we assume that the survival probability at one month $\Delta t = 1/12$ is $S(\Delta t) \approx 0.999$. Then, we use the approximation 
\begin{equation*}
h_0 = h(0) = -\dfrac{S'(0)}{S(0)} \approx -\dfrac{S'(\Delta t)}{S(\Delta t)} \approx  -\dfrac{S(\Delta t) - S(0)}{\Delta t S(\Delta t)} \approx 0.01.
\end{equation*}
To define the initial condition on $q$, we use that treatment does not usually start at the beginning of follow-up. Thus, the response on reducing the hazard function should be small at the beginning of follow-up, and we fix this value at $q_0=10^{-6}$. One could also assign prior distributions to the initial conditions $h_0$ and $q_0$, as long as they are consistent with these values. As a sensitivity analysis, we provide such implementation in the software provided on our GitHub repositories (\url{https://github.com/FJRubio67/ODESurv} and \url{https://github.com/andreschristen/ODESurv}), where we show that very similar results are obtained with both approaches.

\begin{figure}
\begin{center}
\begin{tabular}{c c}
\includegraphics[scale=0.35]{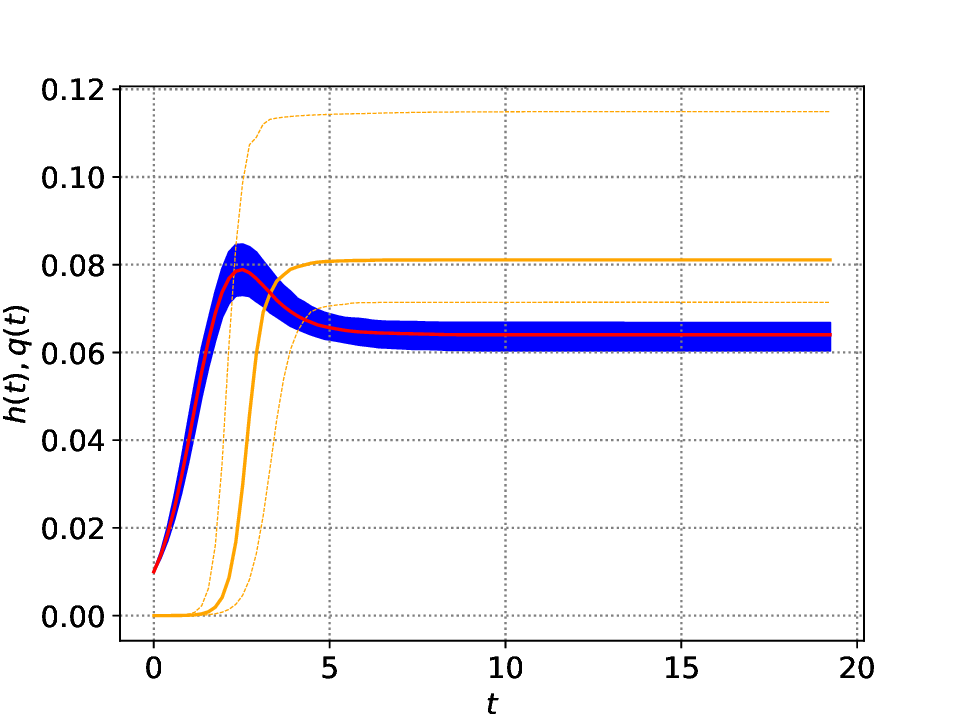} &
\includegraphics[scale=0.35]{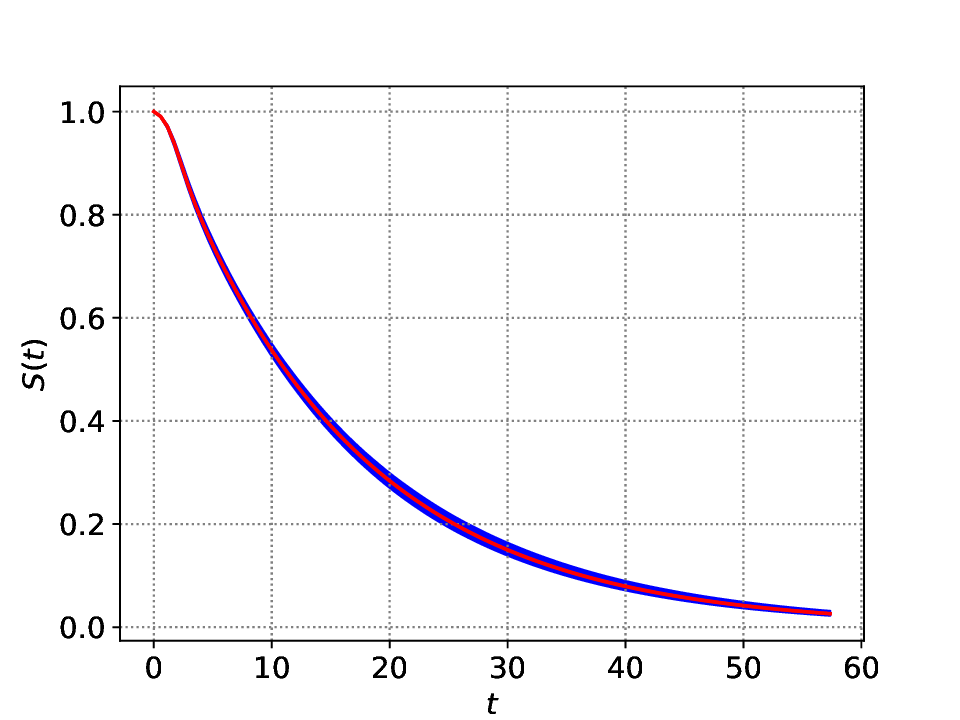} \\
(a) & (b)\\
\includegraphics[scale=0.35]{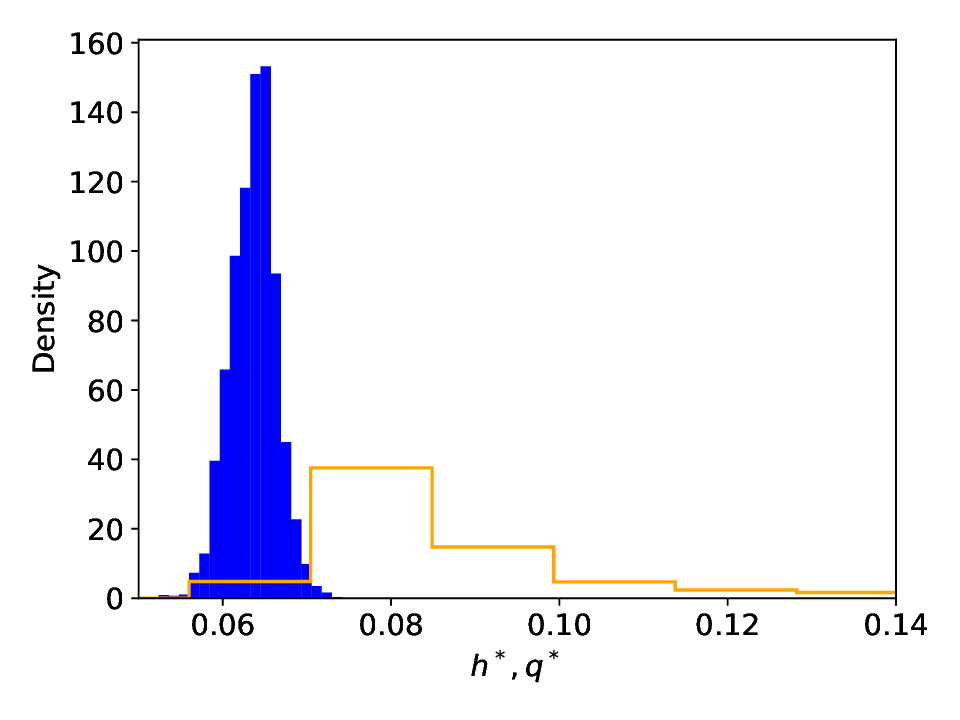} &
\includegraphics[scale=0.265]{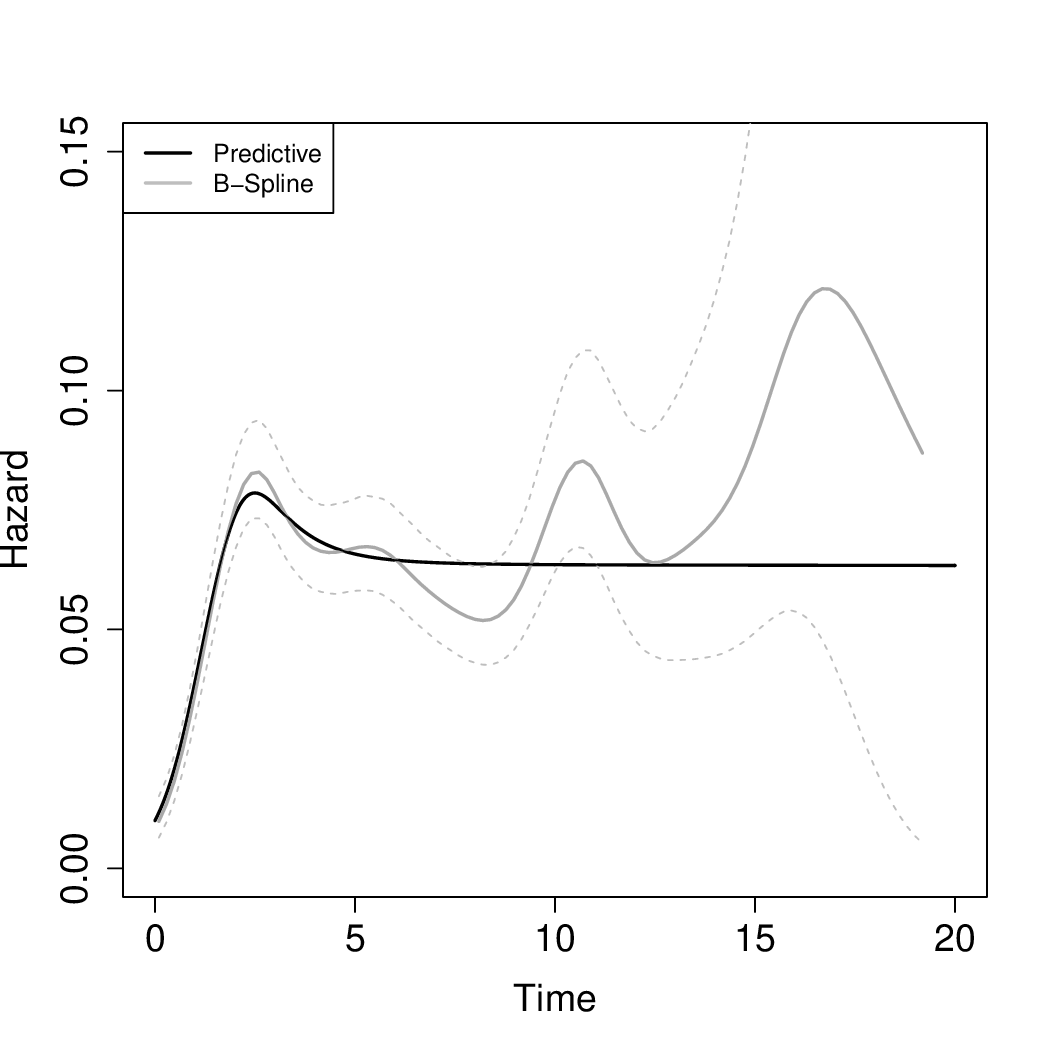} \\
(c) 
\end{tabular}
\end{center}
\caption{\texttt{rotterdam} data: (a) posterior predictive hazard function (red line) and response function (yellow line); (b) survival function (red line) for the Hazard-Response model \eqref{eq:hazardresponse}, 0.1 to 0.9 quantiles and the median; (c) Posterior distribution of $h^*$ (left) and $q^*$ (right); and (d) posterior predictive hazard function (black line) and B-spline estimator (gray line) with $95\%$ confidence interval. The posterior predictive for the ``response'' function $q$ is also included (transparent percentile range) along with the hazard function (a). The system is in equilibrium with probability close to $1$, and $h$ tends to the asymptotic point $h^*$.}
 \label{fig:rott_h_post}
\end{figure}

Figure \ref{fig:rott_h_post} shows the posterior predictive hazard, survival, and ``response'' functions, as well as a comparison against a B-spline estimator of the hazard (using the \texttt{bshazard} R package, with 3 degrees of freedom). 
Note how the hazard function exhibits an increasing behaviour during the first 2 to 3 years and then decreases and stabilises at a constant value. Interestingly, the equilibrium state is inferred and neither $h$ nor $q$ go to zero. That is, a constant asymptotic hazard $h^*$ is predicted for this data set (see Section~\ref{subsec:HT}). This coincides nicely with previous studies on the evolution of the population hazard function associated with breast cancer patients \citep{royston:2002}.  The posterior probability of the equilibrium state may be easily calculated, by checking if $h^* >0, q^*>0$ and $D>0$ at each iteration of the MCMC.  All of our iterations resulted in equilibrium, from an effective sample size of 800.  Moreover, the posterior distribution for $h^*$ is presented in Figure~\ref{fig:rott_h_post}(c). 
Figure~\ref{fig:rott_h_post}(d) illustrates the large variability of the B-splines estimator of the hazard function compared to the hazard-response model. The ``wiggliness'' of the B-splines estimator of the hazard function complicates making precise epidemiological statements about the evolution of the disease.
Finally, we compare the fit of the hazard-response model against the Weibull and Power Generalised Weibull (PGW, 3 parameters) distributions using BIC
\begin{equation*}
BIC = k \log(n) - 2 \ell(\lambda,\kappa,\alpha,\beta) ,  
\end{equation*}
where $k=4$, the total number of unknown parameters, and $n$ is the sample size.
The corresponding BICs are $9638.3, 9572.0$, for the Weibull and PGW distributions, respectively, and $9561.0$ for the hazard-response model. Thus, the BIC clearly favours the proposed ODE model. 

\section{Discussion}\label{sec:discussion}

We proposed a novel methodology for modelling the dynamics of the hazard function in survival analysis via systems of ODEs. This framework is particularly useful for researchers interested in qualitative and quantitative analyses of the evolution of the hazard function over time. 
This modelling approach capitalises on the vast literature on ODEs and systems of ODEs, which have well-understood dynamics and solutions, as well as interpretable parameters. In particular, the models based on autonomous ODEs presented in Section \ref{sec:autonomous} represent a novel and interpretable approach for modelling the dynamics of the hazard function. 
We have presented models with an analytic ODE solution, and consequently their implementation is similar to that of the usual parametric models. We have also shown that numerical ODE solvers can be used in cases where no analytic solution is available, allowing for coupling such models with general-purpose optimisation methods (for likelihood-based inference) as well as MCMC samplers (for Bayesian inference).

The simulation study presented in this work shows it is possible to recover the true parameter values and the shape of the true hazard functions using the proposed modelling approach. This study also provides guidelines about the effect of the sample size and censoring rates on the inference on the parameters, and the increased uncertainty induced by adding more parameters or additional hidden state variables.
The case studies presented here illustrate the use of the proposed modelling approach using real data, as well as the qualitative interpretation of the hazard function in the corresponding context. In particular, Case study II presents an application of the hazard-response model \eqref{eq:hazardresponse}, which allows for an interpretation of the competing processes associated to the mortality hazard and the response, at the population level, potentially associated to clinical interventions and the natural immunological response. 
We conducted model selection using the classical definition of the Bayesian Information Criterion (BIC). Alternatively, one could employ modified versions of BIC tailored for censored samples, or opt for Bayesian methods for model selection, such as the use of Bayes factors or model posterior probabilities.

An important point to consider when choosing an ODE model for the hazard function is the inclusion of a scale parameter. Most classical ODEs already include a scale parameter (potentially under a different parameterisation) to allow for maintaining the same shape under different time scales. However, some textbooks may present ODE formulations without scale parameters, typically to conduct dimensionless analyses. Overall, it is important to keep in mind the need for including scale parameters, as with any other parametric model.

A natural extension of our work consists of using other systems of ODEs to model the hazard function. These include the use of other common population growth models \citep{simpson:2022}, such as Gompertz or Richard's models, or extensions of competition models. For instance, similar to \cite{farkas:1984}, we could incorporate the value of the hazard function in the past into the effect on the response (\textit{i.e.} the response depends on the history and evolution of the hazard) as follows:
\begin{eqnarray*}
\begin{cases}
h'(t)  =  \lambda h(t) \left(1 - \dfrac{h(t)}{\kappa}\right) - \alpha q(t) h(t), & h(0) = h_0 \\
q'(t) =  \beta q(t) \left( 1- \frac{q(t)}{\kappa} \right) -\alpha q(t) \int_{-\infty}^t h(\tau) g(t-\tau)d\tau  ,  &  q(0) = q_0 \\ 
H'(t)  =  h(t), & H(0) = 0,
\end{cases}
\label{eq:hazardresponse_ext}
\end{eqnarray*}
where $g(t)=\eta \exp(-\eta t)$, $\eta>0$. This represents a Volterra-type integro-differential equation. Since the integral in the previous equation represents a convolution, the integro-differential system can be rewritten in terms of a system of ODEs with an additional state, which avoids the need for numerical integration and allows for applying the methodology proposed in this work,
\begin{eqnarray*}
\begin{cases}
h'(t)  =  \lambda h(t) \left(1 - \dfrac{h(t)}{\kappa}\right) - \alpha q(t) h(t), & h(0) = h_0 \\
q'(t) =  \beta q(t) \left( 1- \frac{q(t)}{\kappa} \right) -\alpha q(t) \tilde{q}(t)  ,  &  q(0) = q_0 \\
\tilde{q}'(t) = \eta[ h(t) - \tilde{q}(t)]  ,  &  \tilde{q}(0) = \tilde{q}_0 \\
H'(t)  =  h(t), & H(0) = 0,
\end{cases}
\label{eq:hazardresponse_ext2}
\end{eqnarray*}
This illustrates the variety and richness of models one can obtain by adding hidden states, as well as the interpretability of such additional states. 

We have focused on the case without covariates, as we aimed at providing a deeper analysis of the proposed modelling strategy. However, our modelling strategy allows for seamlessly introducing covariates using standard approaches. For instance, one could use the solution to \eqref{eq:general_ode} as a baseline hazard in any hazard-based regression model. This includes the proportional hazards, accelerated failure time, accelerated hazards, or general/extended hazards models (see \citealp{rubio:2019} for a review on classical hazard structures in survival analysis). Another approach, that takes advantage of the interpretability of the model parameters, consists of modelling them using a transformed linear or additive predictor. These strategies will be explored in future research. The proposed methodology can be extended to other models of interest in survival analysis such as cure models, spatial survival models, relative survival models \citep{rubio:2019}, and etcetera.

\clearpage

\section*{Appendix}\label{sec:appendix}
\subsection*{Marginal posteriors for real data applications}

\begin{figure}[htp]
\begin{center}
\begin{tabular}{c c c}
\includegraphics[scale = 0.3]{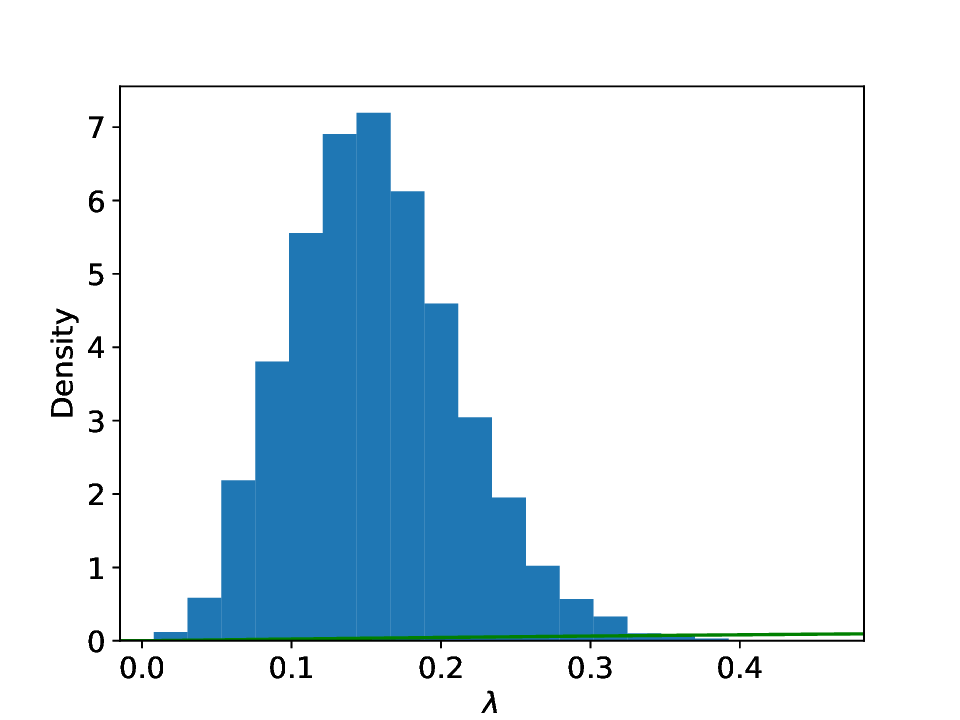} &
\includegraphics[scale = 0.3]{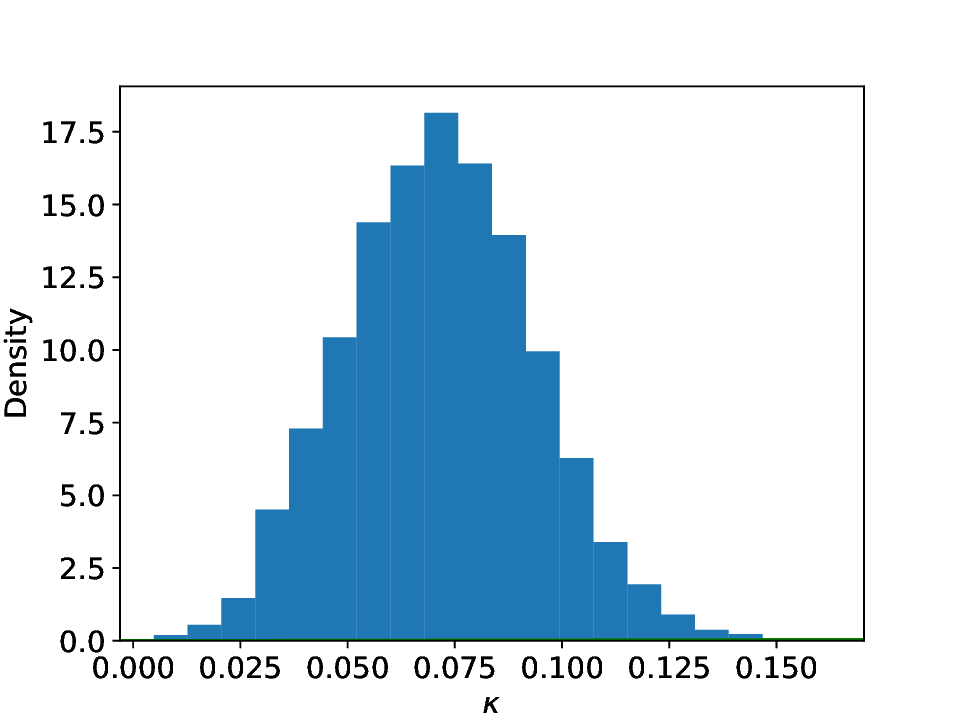} &
\includegraphics[scale = 0.3]{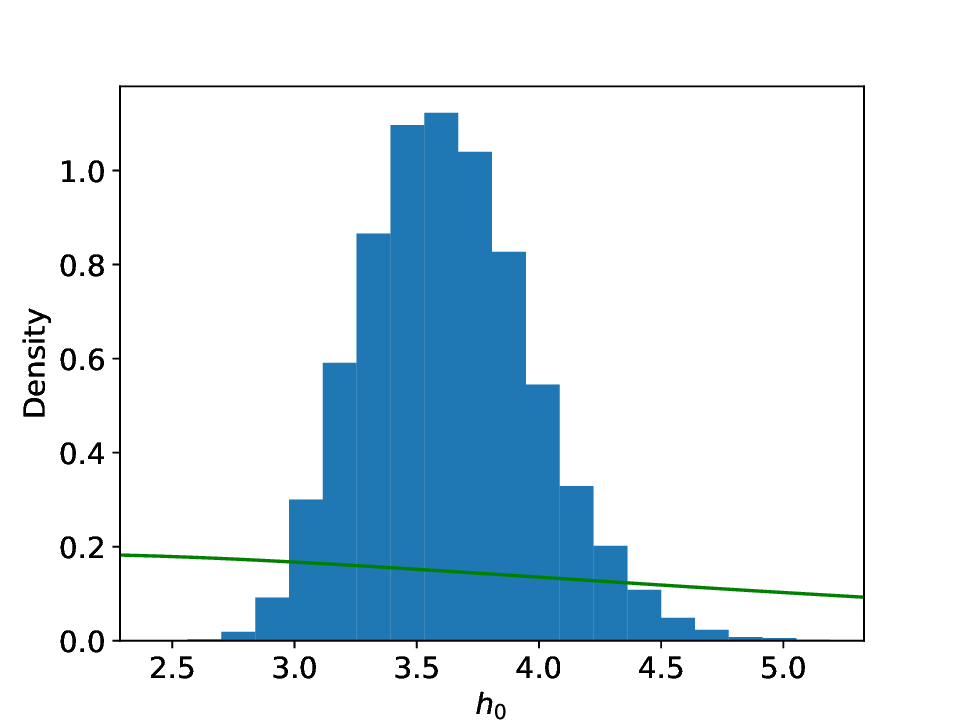} \\
(a) & (b) & (c) \\
\end{tabular}    
 \caption{\texttt{LeukSurv} data: posterior distributions of the parameters of the logistic ODE model. (a) posterior of $\lambda$, (b) posterior of $\kappa$, and (c) posterior of $h_0$.}
 \label{fig:logisODE}
\end{center}
\end{figure}

\begin{figure}
\begin{center}
\begin{tabular}{c c}
\includegraphics[scale=0.5]{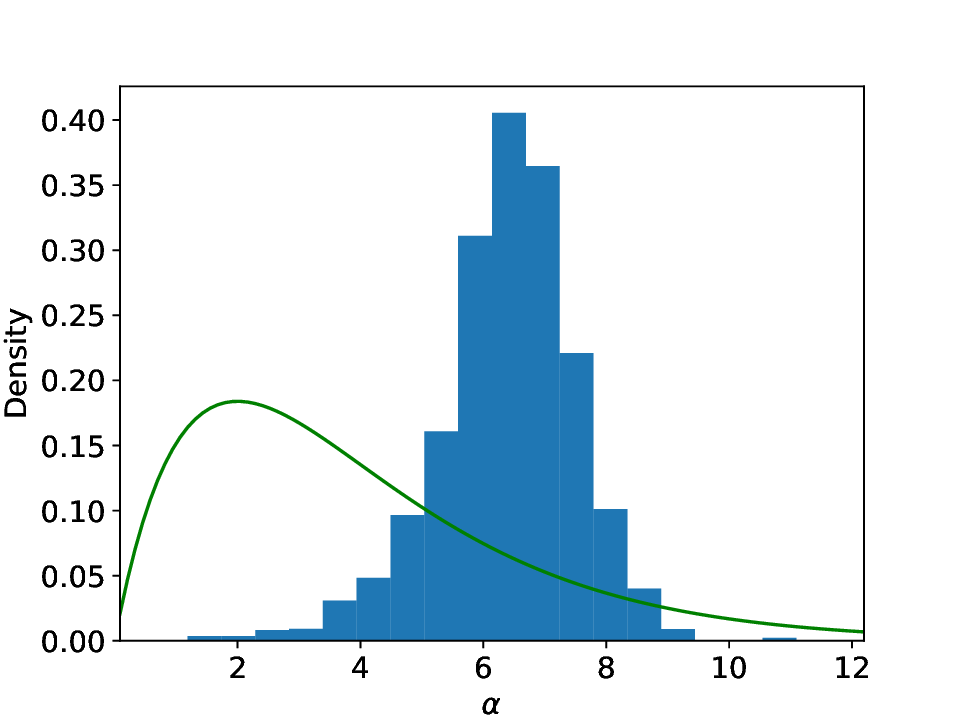} &
\includegraphics[scale=0.5]{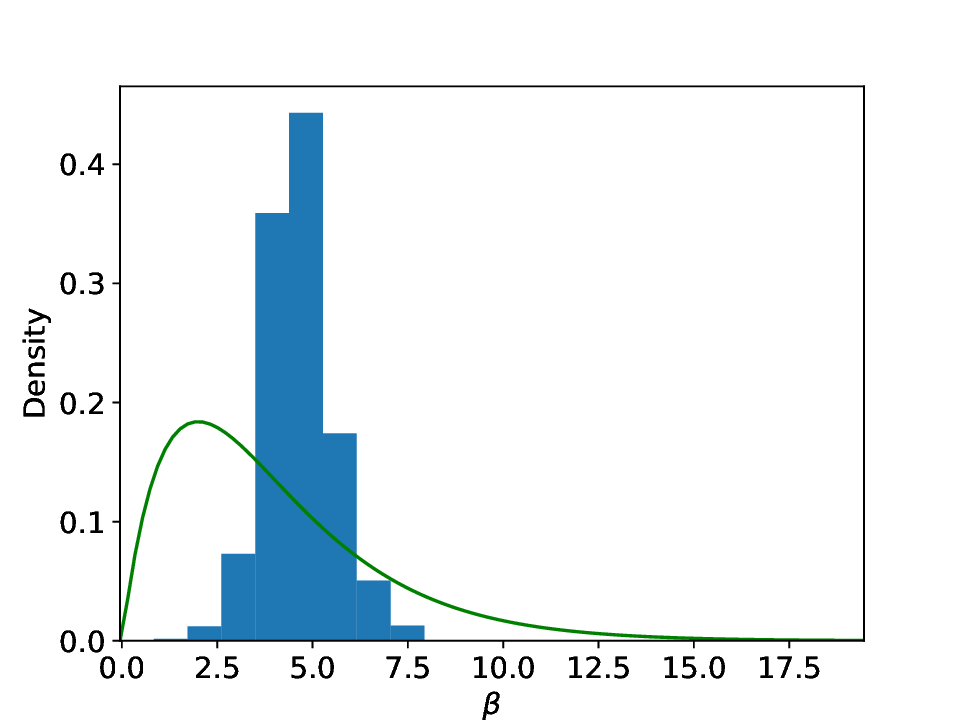} \\
(a) & (b) \\
\includegraphics[scale=0.5]{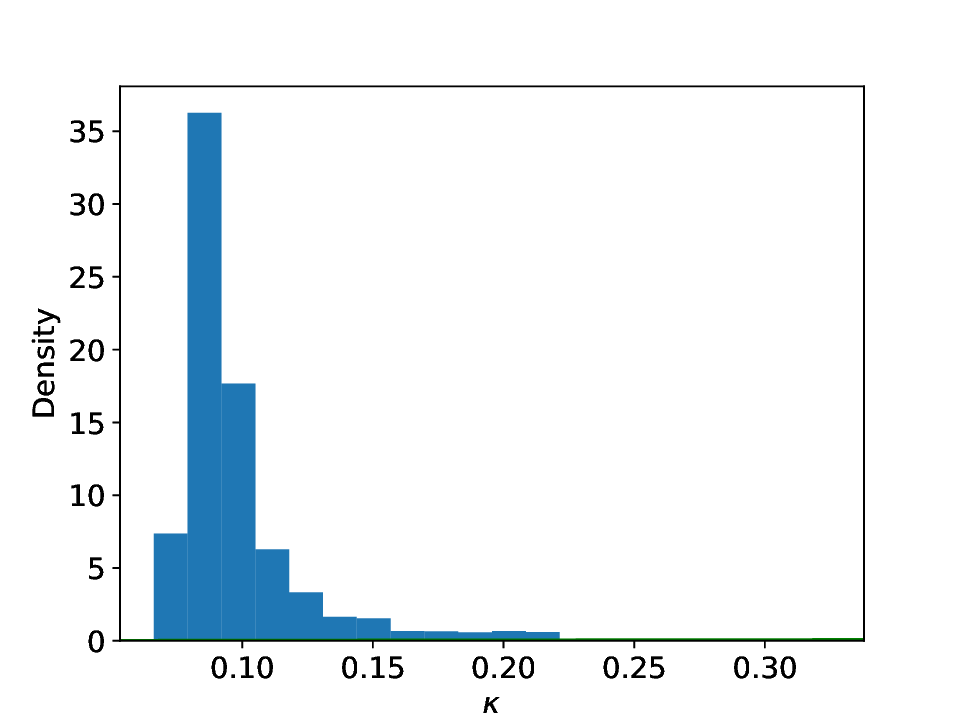} &
\includegraphics[scale=0.5]{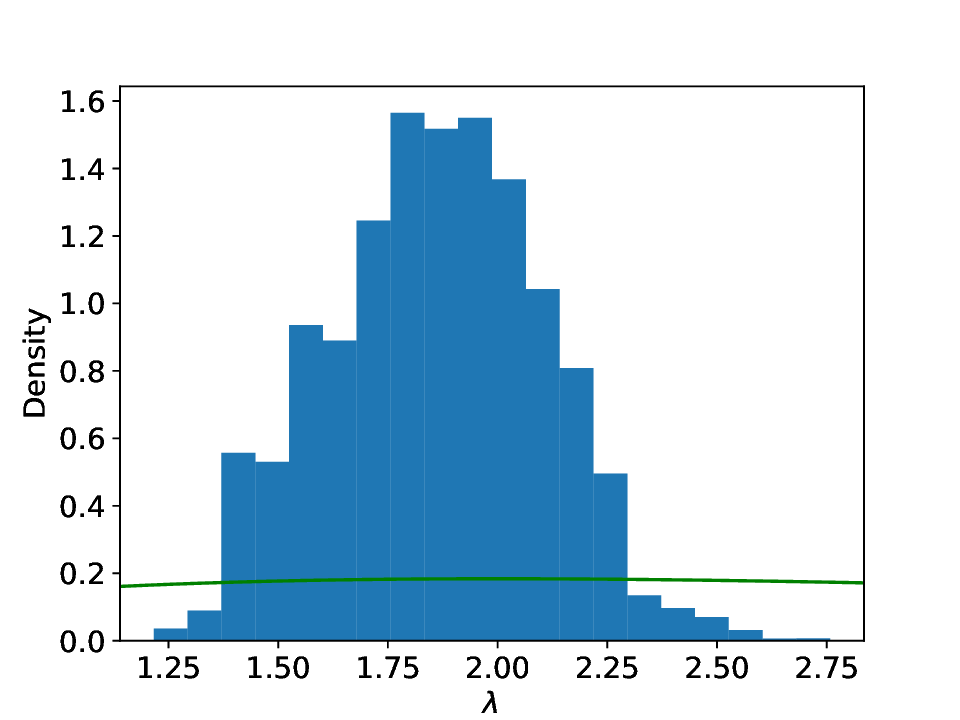} \\
(c) & (d)
\end{tabular}
\end{center}
\caption{\texttt{rotterdam} data: Marginal posteriors for the Hazard-Response hazard function. The relevant part of the prior density is shown in green.}
 \label{fig:rott_post_samp}
\end{figure}

\subsection*{Jacobian for the Hazard-Response system of ODEs}
The Jacobian for the Hazard-Response system of ODEs \eqref{eq:hazardresponse} is:
\begin{eqnarray*}
    J_{HT}(t,h,q,H,\btheta) = \begin{pmatrix}
\lambda - \dfrac{2\lambda h}{\kappa}  - \alpha q & -\alpha h & 0\\
- \alpha q & \beta - \dfrac{2\beta q}{\kappa} -\alpha h & 0 \\
1 & 0 & 0 
\end{pmatrix},
\end{eqnarray*}
where $\btheta = (\lambda,\kappa,\alpha,\beta)$ are the model parameters.

\subsection*{Logarithmic formulation of the hazard-response model}

Define $\tilde{h}(t) = \log h(t)$ and $\tilde{q}(t) = \log q(t)$. Then, the hazard-response model \eqref{eq:hazardresponse} becomes
\begin{eqnarray*}
\begin{cases}
\tilde{h}'(t)  =  \lambda  \left(1 - \dfrac{\exp\left\{\tilde{h}(t) \right\}}{\kappa}\right) - \alpha \exp\left\{ \tilde{q}(t)\right\}, & \tilde{h}(0) = \log\left\{h_0\right\} \\
\tilde{q}'(t) =  \beta  \left( 1- \dfrac{\exp\left\{\tilde{q}(t) \right\}}{\kappa} \right) -\alpha \exp\left\{\tilde{h}(t) \right\}  ,  &  \tilde{q}(0) = \log\left\{q_0\right\} \\ 
H'(t)  = \exp\left\{\tilde{h}(t) \right\}, & H(0) = 0.
\end{cases}
\label{eq:loghazardresponse}
\end{eqnarray*}
The corresponding Jacobian  is:
\begin{eqnarray*}
    J_{HT}(t,\tilde{h},\tilde{q},H,\btheta) = \begin{pmatrix}
 - \dfrac{\lambda}{\kappa}\exp\left\{\tilde{h}(t) \right\} & -\alpha \exp\left\{ \tilde{q}(t)\right\} & 0\\
- \alpha \exp\left\{\tilde{h}(t)\right\} & - \dfrac{\beta}{\kappa}\exp\left\{\tilde{q}(t) \right\}   & 0 \\
\exp\left\{\tilde{h}(t) \right\} & 0 & 0 
\end{pmatrix},
\end{eqnarray*}
where $\btheta = (\lambda,\kappa,\alpha,\beta)$ are the model parameters.
\subsection*{A note on the posterior predictive hazard}

Given $\pi( \btheta \mid \text{data})$ the posterior density for the parameter $\btheta$ of a hazard function
$h_{\btheta}(t)$, one may calculate the posterior distribution of the hazard function, since at any $t$, $h(t \mid \btheta) = \frac{f (t \mid \btheta) }{S (t \mid \btheta) } \in \mathbb{R}^+$ and this is simply a function of the random variable $\btheta$.  From a MCMC sample of the posterior, $\btheta^{(1)}, \ldots , \btheta^{(M)}$, one obtains a sample of the posterior hazard from
$h(t \mid \btheta^{(1)}), \ldots , h(t \mid \btheta^{(M)})$. 

On the other hand, we may calculate the predictive distribution of a hypothetical future $n+1$ patient
$$
f_{t_{n+1}} ( t \mid \text{data}) = \int_{\Theta} f(t \mid \btheta) \pi( \btheta \mid \text{data}) d\btheta =
\int_{\Theta} h (t \mid \btheta) \exp(-H(t \mid \btheta)) \pi( \btheta \mid \text{data}) d\btheta.
$$
From this \textit{fixed} posterior predictive density we may calculate its hazard function
$$
h_{t_{n+1}} ( t \mid \text{data}) = \frac{f_{t_{n+1}} ( t \mid \text{data})}{1 - \int_0^t f_{t_{n+1}} ( r \mid \text{data}) dr} .
$$
This is the posterior predictive hazard and is the hazard to be analysed, if we wish to discuss the prognosis of a future patient, whereas the former represents the posterior uncertainty we have about the hazard in the population.  

Unfortunately, besides the simplest cases, calculating $h_{t_{n+1}} ( t \mid \text{data})$ in closed-form is not possible.  
However, we can obtain a Monte Carlo approximation using a posterior sample of the parameters, as follows
\begin{equation*}
f_{t_{n+1}}(t\mid \text{data})   \approx \dfrac{1}{M} \sum_{j=1}^M h (t \mid \btheta^{(j)}) \exp\left\{ - H (t \mid \btheta^{(j)})  \right\}
\end{equation*}
and
\begin{equation*}
1 - \int_0^t f_{t_{n+1}} ( r \mid \text{data}) dr   \approx \dfrac{1}{M} \sum_{j=1}^M \exp\left\{ - H (t \mid \btheta^{(j)})  \right\}.
\end{equation*}
Consequently
$$
h_{t_{n+1}} ( t \mid \text{data}) \approx \dfrac{ \dfrac{1}{M} \sum_{j=1}^M h (t \mid \btheta^{(j)}) \exp\left\{ - H (t \mid \btheta^{(j)}) \right\}}{ \dfrac{1}{M} \sum_{j=1}^M \exp\left\{ - H (t \mid \btheta^{(j)})  \right\}} .
$$

Indeed, and by the same argument, the predictive survival function may be approximated by
$$
S_{t_{n+1}} ( t \mid \text{data}) \approx 
\dfrac{1}{M} \sum_{j=1}^M \exp\left\{ - H (t \mid \btheta^{(j)})  \right\} .
$$

\pagebreak

\subsection*{Simulation study results: Logistic ODE hazard model}

\begin{table}[ht]
\centering
\begin{tabular}{| c c c c c c |}
  \hline
  Parameter & Mean & Median & SD & RMSE & coverage \\
  \hline
  \multicolumn{6}{|c|}{$n=250$}   \\
$\lambda$ (0.5) & 0.597 & 0.582 & 0.138 & 0.163 & 0.916 \\ 
  $\kappa$ (0.05) & 0.053 & 0.052 & 0.006 & 0.007 & 0.952 \\ 
  $h_0$ (3.5) & 4.349 & 4.137 & 1.372 & 1.436 & 0.948 \\ 
    \multicolumn{6}{|c|}{$n=500$}  \\
$\lambda$ (0.5) & 0.545 & 0.538 & 0.089 & 0.100 & 0.932 \\ 
  $\kappa$ (0.05) & 0.051 & 0.051 & 0.004 & 0.004 & 0.936 \\ 
  $h_0$ (3.5) & 3.976 & 3.863 & 0.914 & 0.983 & 0.932 \\
    \multicolumn{6}{|c|}{$n=1000$}  \\
$\lambda$ (0.5) & 0.527 & 0.523 & 0.061 & 0.070 & 0.924 \\ 
  $\kappa$ (0.05) & 0.051 & 0.051 & 0.003 & 0.003 & 0.944 \\ 
  $h_0$ (3.5) & 3.754 & 3.697 & 0.617 & 0.656 & 0.932 \\ 
    \multicolumn{6}{|c|}{$n=5000$}  \\
$\lambda$ (0.5) & 0.506 & 0.505 & 0.026 & 0.027 & 0.936 \\ 
  $\kappa$ (0.05) & 0.050 & 0.050 & 0.001 & 0.001 & 0.932 \\ 
  $h_0$ (3.5) & 3.530 & 3.519 & 0.259 & 0.265 & 0.944 \\    
   \hline
\end{tabular}
\caption{Simulation scenario 1 with $25\%$ censoring rate. Average posterior mean, average posterior median, average posterior standard deviation, average RMSE with respect to the posterior mean, and coverage of the $95\%$ credible intervals.}
\label{tab:S1C25}
\end{table}

\begin{table}[ht]
\centering
\begin{tabular}{| c c c c c c |}
  \hline
  Parameter & Mean & Median & SD & RMSE & coverage \\
  \hline
  \multicolumn{6}{|c|}{$n=250$}   \\
$\lambda$ (0.5) & 0.704 & 0.679 & 0.237 & 0.298 & 0.920 \\ 
  $\kappa$ (0.05) & 0.059 & 0.059 & 0.013 & 0.014 & 0.936 \\ 
  $h_0$ (3.5) & 4.448 & 4.226 & 1.431 & 1.501 & 0.936 \\ 
    \multicolumn{6}{|c|}{$n=500$}  \\
$\lambda$ (0.5) & 0.600 & 0.589 & 0.156 & 0.182 & 0.912 \\ 
  $\kappa$ (0.05) & 0.054 & 0.054 & 0.010 & 0.011 & 0.920 \\ 
  $h_0$ (3.5) & 4.032 & 3.915 & 0.946 & 1.017 & 0.940 \\
    \multicolumn{6}{|c|}{$n=1000$}  \\
$\lambda$ (0.5) & 0.558 & 0.554 & 0.109 & 0.122 & 0.924 \\ 
  $\kappa$ (0.05) & 0.052 & 0.053 & 0.007 & 0.007 & 0.944 \\ 
  $h_0$ (3.5) & 3.789 & 3.731 & 0.637 & 0.680 & 0.952 \\
    \multicolumn{6}{|c|}{$n=5000$}  \\
$\lambda$ (0.5) & 0.510 & 0.509 & 0.047 & 0.045 & 0.968 \\ 
  $\kappa$ (0.05) & 0.050 & 0.050 & 0.003 & 0.003 & 0.952 \\ 
  $h_0$ (3.5) & 3.534 & 3.522 & 0.264 & 0.268 & 0.960 \\ 
   \hline
\end{tabular}
\caption{Simulation scenario 1 with $50\%$ censoring rate. Average posterior mean, average posterior median, average posterior standard deviation, average RMSE with respect to the posterior mean, and coverage of the $95\%$ credible intervals.}
\label{tab:S1C50}
\end{table}

\pagebreak
\subsection*{Simulation study results: hazard-response model}

\begin{table}[ht]
\centering
\begin{tabular}{| c c c c c c |}
  \hline
  Parameter & Mean & Median & SD & RMSE & coverage \\
  \hline
  \multicolumn{6}{|c|}{$n=250$}   \\
$\lambda$ (1.8) & 2.446 & 1.755 & 0.850 & 1.041 & 0.965 \\ 
  $\kappa$ (0.1) & 0.141 & 0.082 & 0.076 & 0.088 & 0.985 \\ 
  $\alpha$ (6) & 4.224 & 3.169 & 1.812 & 1.925 & 1.000 \\ 
  $\beta$ (4.8) & 4.824 & 3.567 & 2.171 & 0.674 & 1.000 \\  
    \multicolumn{6}{|c|}{$n=500$}  \\
$\lambda$ (1.8) & 2.112 & 1.598 & 0.494 & 0.616 & 0.955 \\ 
  $\kappa$ (0.1) & 0.162 & 0.105 & 0.074 & 0.156 & 0.970 \\ 
  $\alpha$ (6) & 4.297 & 3.334 & 1.615 & 1.954 & 0.965 \\ 
  $\beta$ (4.8) & 4.862 & 3.683 & 1.904 & 0.869 & 0.985 \\ 
    \multicolumn{6}{|c|}{$n=1000$}  \\
$\lambda$ (1.8) & 2.014 & 1.571 & 0.321 & 0.412 & 0.960 \\ 
  $\kappa$ (0.1) & 0.130 & 0.088 & 0.046 & 0.101 & 0.975 \\ 
  $\alpha$ (6) & 4.665 & 3.755 & 1.426 & 1.646 & 0.975 \\ 
  $\beta$ (4.8) & 4.712 & 3.637 & 1.601 & 0.767 & 0.990 \\ 
    \multicolumn{6}{|c|}{$n=5000$}  \\
$\lambda$ (1.8) & 1.832 & 1.460 & 0.154 & 0.183 & 0.965 \\ 
  $\kappa$ (0.1) & 0.119 & 0.088 & 0.027 & 0.053 & 0.980 \\ 
  $\alpha$ (6) & 5.303 & 4.326 & 0.745 & 1.053 & 0.950 \\ 
  $\beta$ (4.8) & 4.901 & 3.880 & 0.824 & 0.799 & 0.965 \\ 
   \hline
\end{tabular}
\caption{Simulation scenario 2 with $25\%$ censoring rate. Average posterior mean, average posterior median, average posterior standard deviation, average RMSE with respect to the posterior mean, and coverage of the $95\%$ credible intervals.}
\label{tab:S2C25}
\end{table}

\begin{table}[ht]
\centering
\begin{tabular}{| c c c c c c |}
  \hline
  Parameter & Mean & Median & SD & RMSE & coverage \\
  \hline
  \multicolumn{6}{|c|}{$n=250$}   \\
$\lambda$ (1.8) & 2.374 & 1.698 & 0.839 & 0.996 & 0.975 \\ 
  $\kappa$ (0.1) & 0.211 & 0.102 & 0.158 & 0.212 & 0.995 \\ 
  $\alpha$ (6) & 3.989 & 2.921 & 1.848 & 2.130 & 0.990 \\ 
  $\beta$ (4.8) & 5.166 & 3.878 & 2.268 & 0.934 & 1.000 \\   
    \multicolumn{6}{|c|}{$n=500$}  \\
$\lambda$ (1.8) & 2.098 & 1.580 & 0.510 & 0.594 & 0.985 \\ 
  $\kappa$ (0.1) & 0.183 & 0.100 & 0.112 & 0.176 & 0.990 \\ 
  $\alpha$ (6) & 4.247 & 3.248 & 1.728 & 1.958 & 0.985 \\ 
  $\beta$ (4.8) & 5.005 & 3.814 & 1.986 & 0.756 & 1.000 \\
    \multicolumn{6}{|c|}{$n=1000$}  \\
 $\lambda$ (1.8) & 1.999 & 1.559 & 0.328 & 0.399 & 0.970 \\ 
  $\kappa$ (0.1) & 0.147 & 0.093 & 0.069 & 0.134 & 0.990 \\ 
  $\alpha$ (6) & 4.564 & 3.643 & 1.549 & 1.704 & 0.980 \\ 
  $\beta$ (4.8) & 4.863 & 3.735 & 1.694 & 0.747 & 1.000 \\ 
    \multicolumn{6}{|c|}{$n=5000$}  \\
$\lambda$ (1.8) & 1.824 & 1.454 & 0.157 & 0.188 & 0.950 \\ 
  $\kappa$ (0.1) & 0.127 & 0.090 & 0.035 & 0.072 & 0.960 \\ 
  $\alpha$ (6) & 5.204 & 4.261 & 0.849 & 1.142 & 0.975 \\ 
  $\beta$ (4.8) & 4.963 & 3.926 & 0.856 & 0.847 & 0.955 \\ 
   \hline
\end{tabular}
\caption{Simulation scenario 2 with $50\%$ censoring rate. Average posterior mean, average posterior median, average posterior standard deviation, average RMSE with respect to the posterior mean, and coverage of the $95\%$ credible intervals.}
\label{tab:S2C50}
\end{table}

\begin{figure}[ht]
\centering
\begin{tabular}{c c c c}
\multicolumn{4}{c}{$25\%$ censoring}\\
    \includegraphics[width = 3.5cm, height = 3cm]{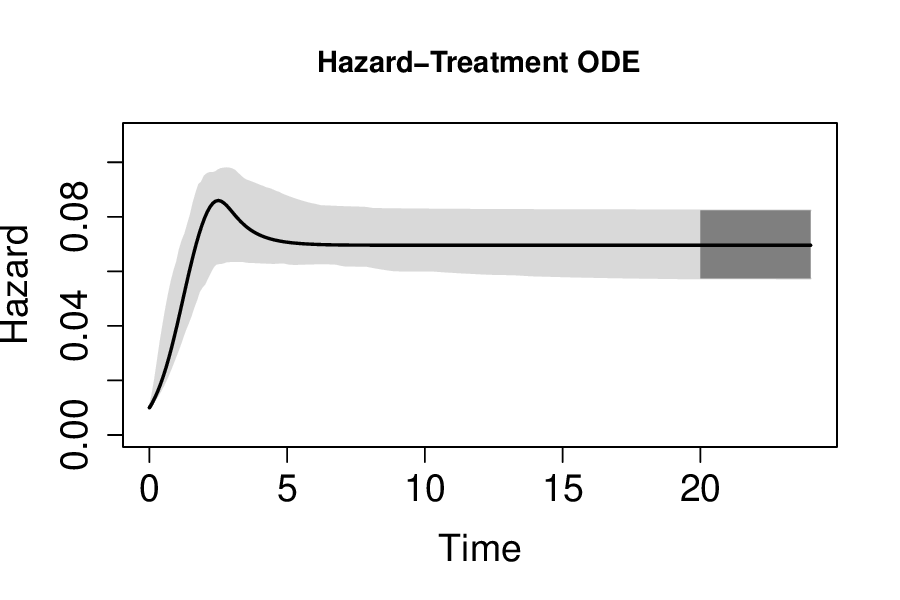} &
    \includegraphics[width = 3.5cm, height = 3cm]{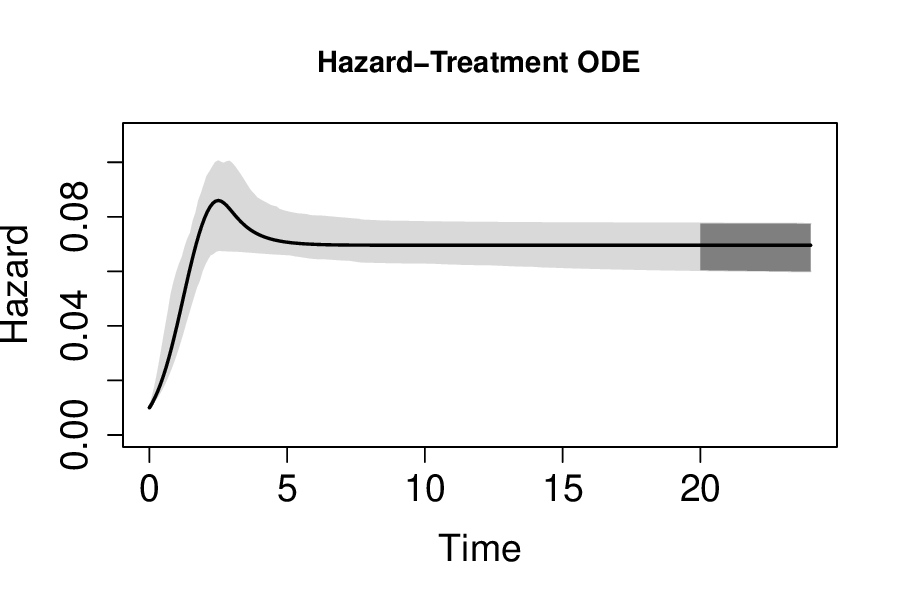} &
    \includegraphics[width = 3.5cm, height = 3cm]{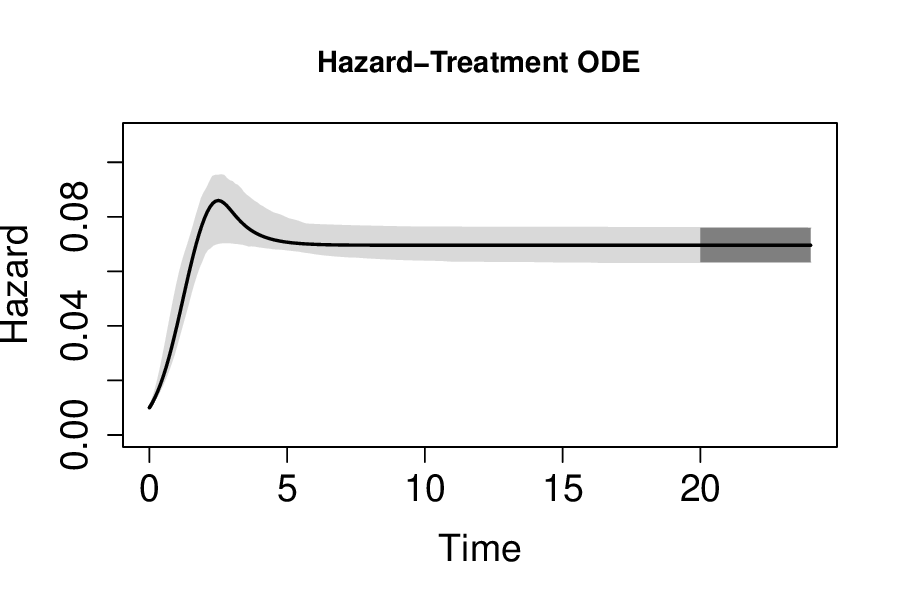} &
      \includegraphics[width = 3.5cm, height = 3cm]{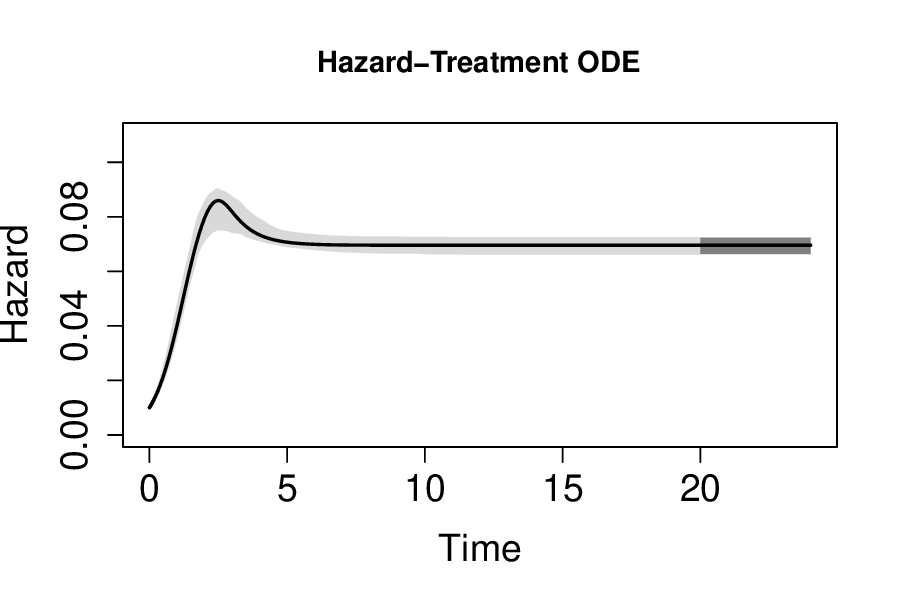}   \\
    (a) $n=250$ & (b) $n=500$ & (c) $n=1000$ & (d) $n=5000$\\
    \multicolumn{4}{c}{$50\%$ censoring}\\
     \includegraphics[width = 3.5cm, height = 3cm]{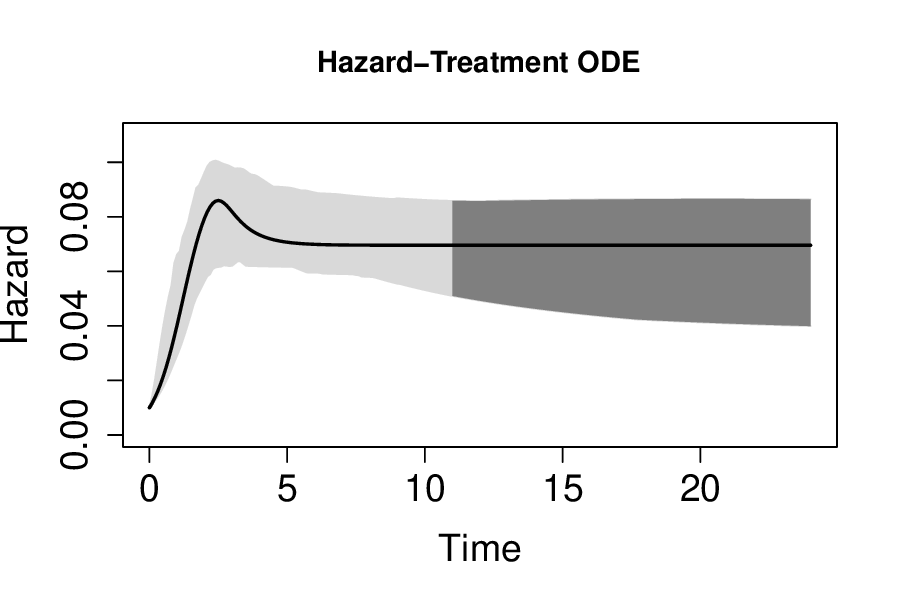} &
    \includegraphics[width = 3.5cm, height = 3cm]{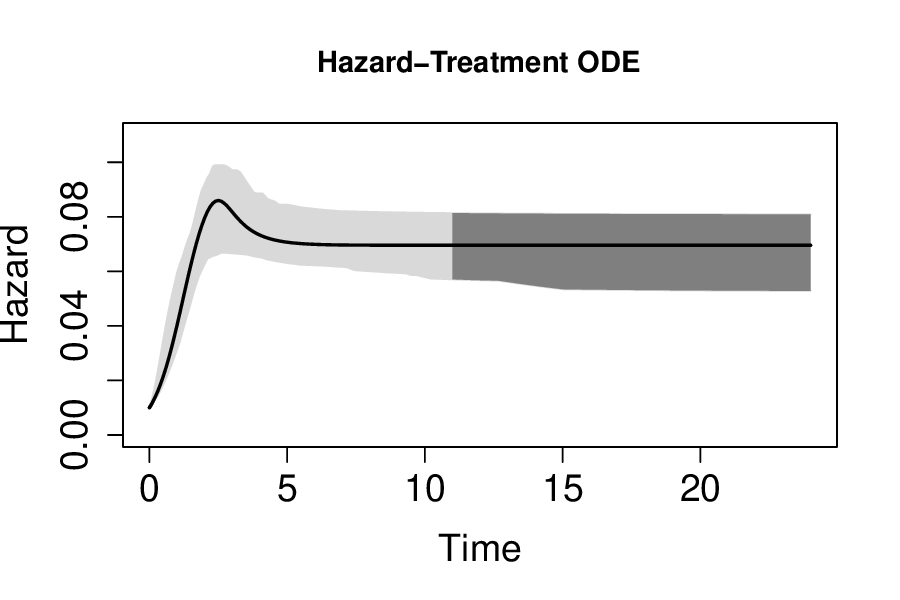} &
    \includegraphics[width = 3.5cm, height = 3cm]{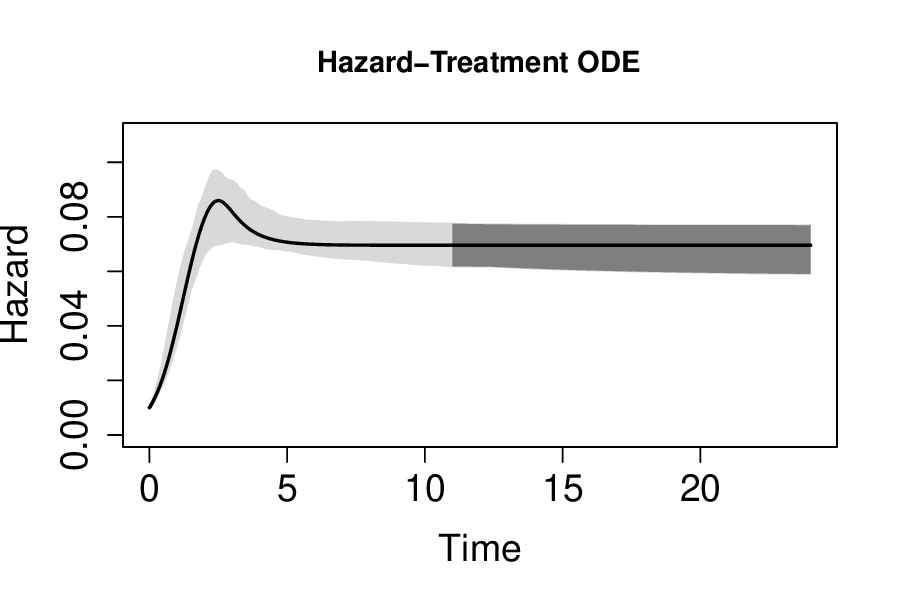} &
     \includegraphics[width = 3.5cm, height = 3cm]{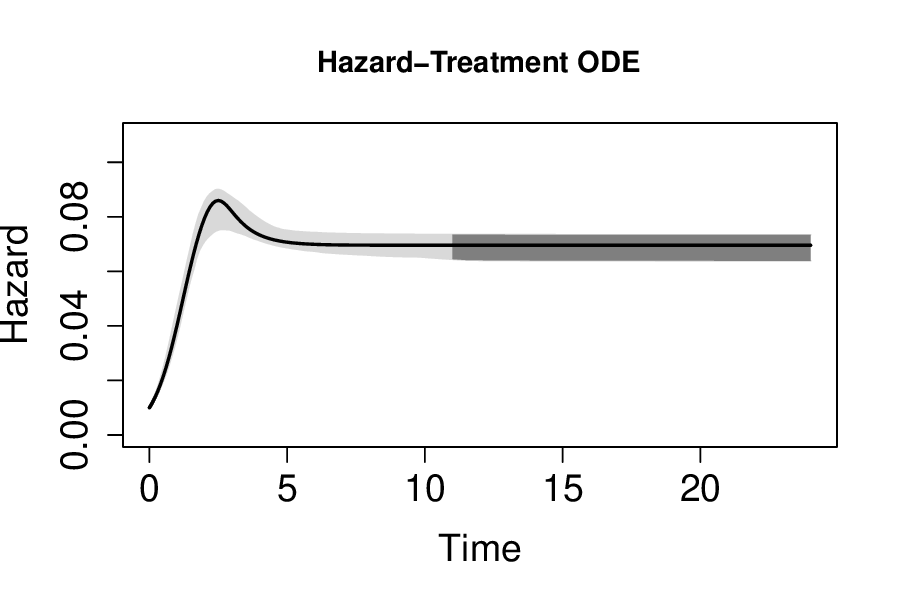}    \\
    (e) $n=250$ & (f) $n=500$& (g) $n=1000$ & (h) $n=5000$
    \end{tabular}
    \caption{Simulation scenario 2: Predictive hazards and $95\%$ predictive intervals. The dark gray area indicates the censoring point.}
    \label{fig:predhazS3}
\end{figure}

\begin{figure}[ht]
\centering
\begin{tabular}{c c}
\multicolumn{2}{c}{$25\%$ censoring}\\
    \includegraphics[width = 6cm, height = 4cm]{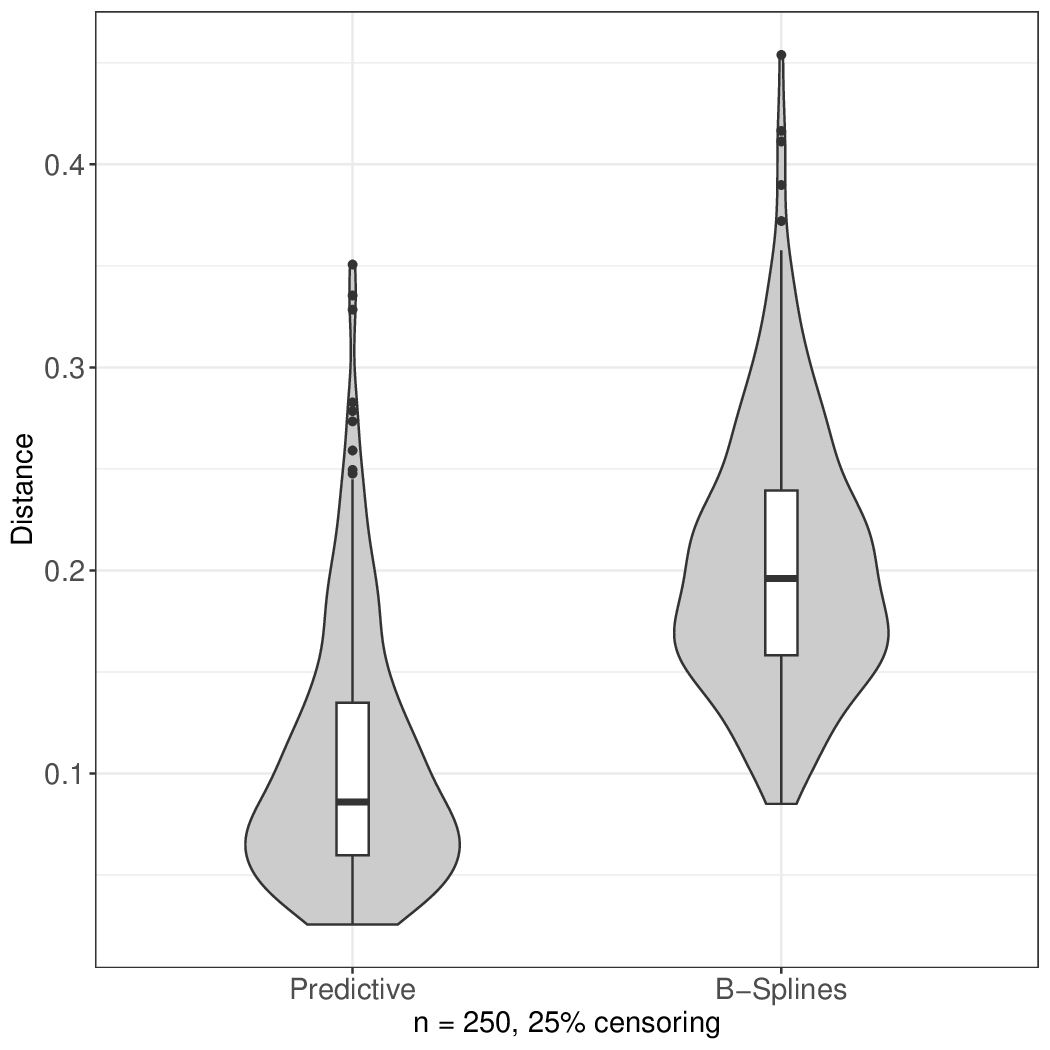} &
    \includegraphics[width = 6cm, height = 4cm]{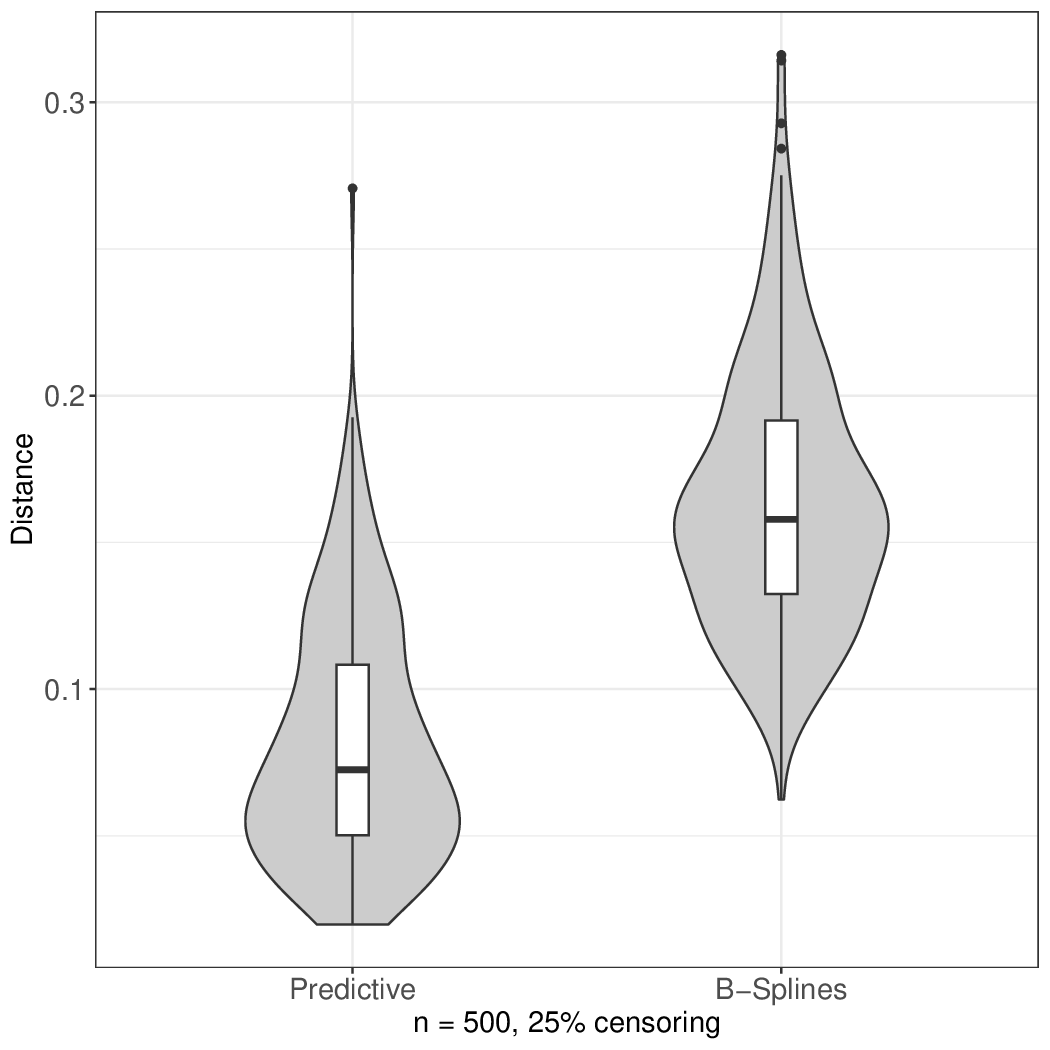} \\
    (a) $n=250$ & (b) $n=500$ \\
    \includegraphics[width = 6cm, height = 4cm]{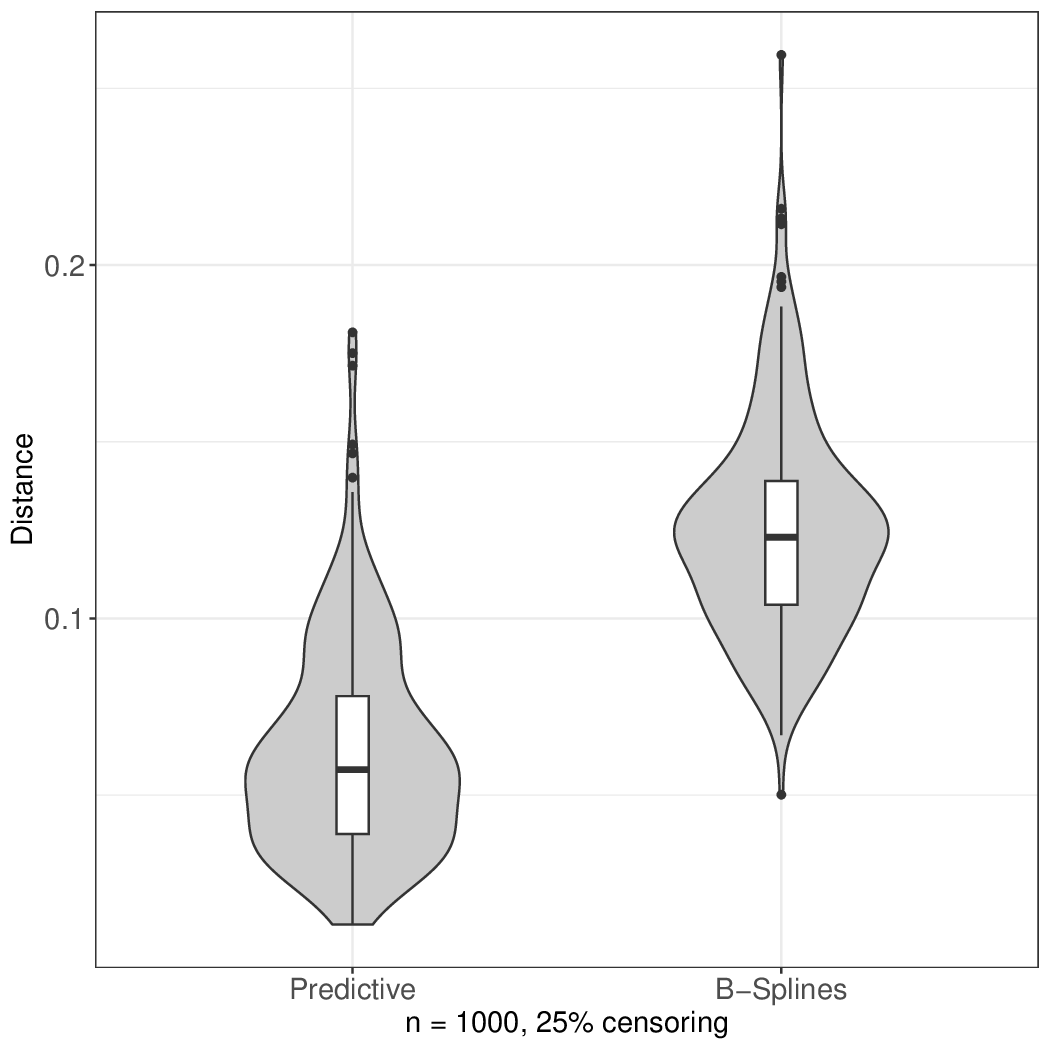} &
      \includegraphics[width = 6cm, height = 4cm]{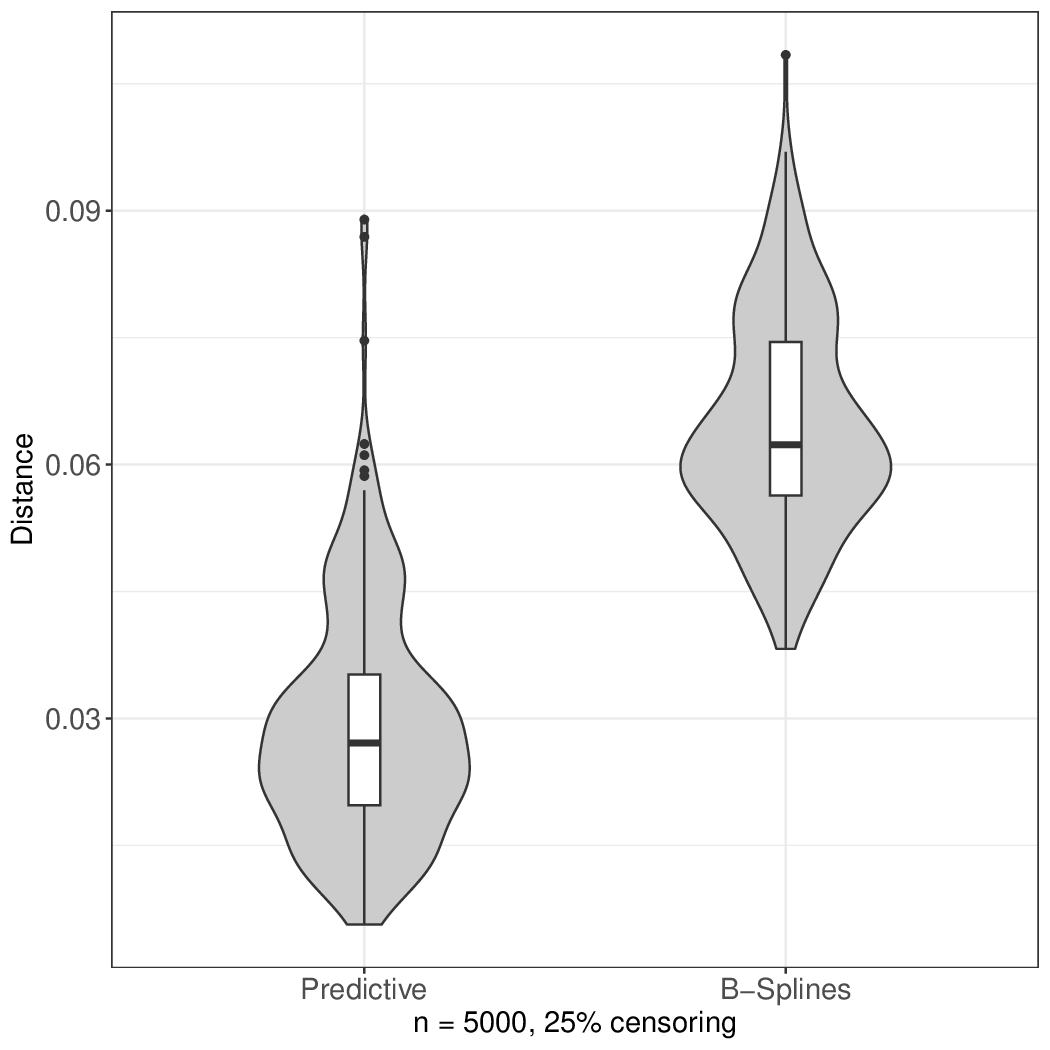}   \\
    (c) $n=1000$ & (d) $n=5000$
    \end{tabular}
    \caption{Simulation scenario 2, $25\%$ censoring: Distance between the posterior predictive hazards and the B-Spline estimator to the true generating model.}
    \label{fig:dist25}
\end{figure}

\begin{figure}[ht]
\centering
\begin{tabular}{c c}
    \multicolumn{2}{c}{$50\%$ censoring}\\
     \includegraphics[width = 6cm, height = 4cm]{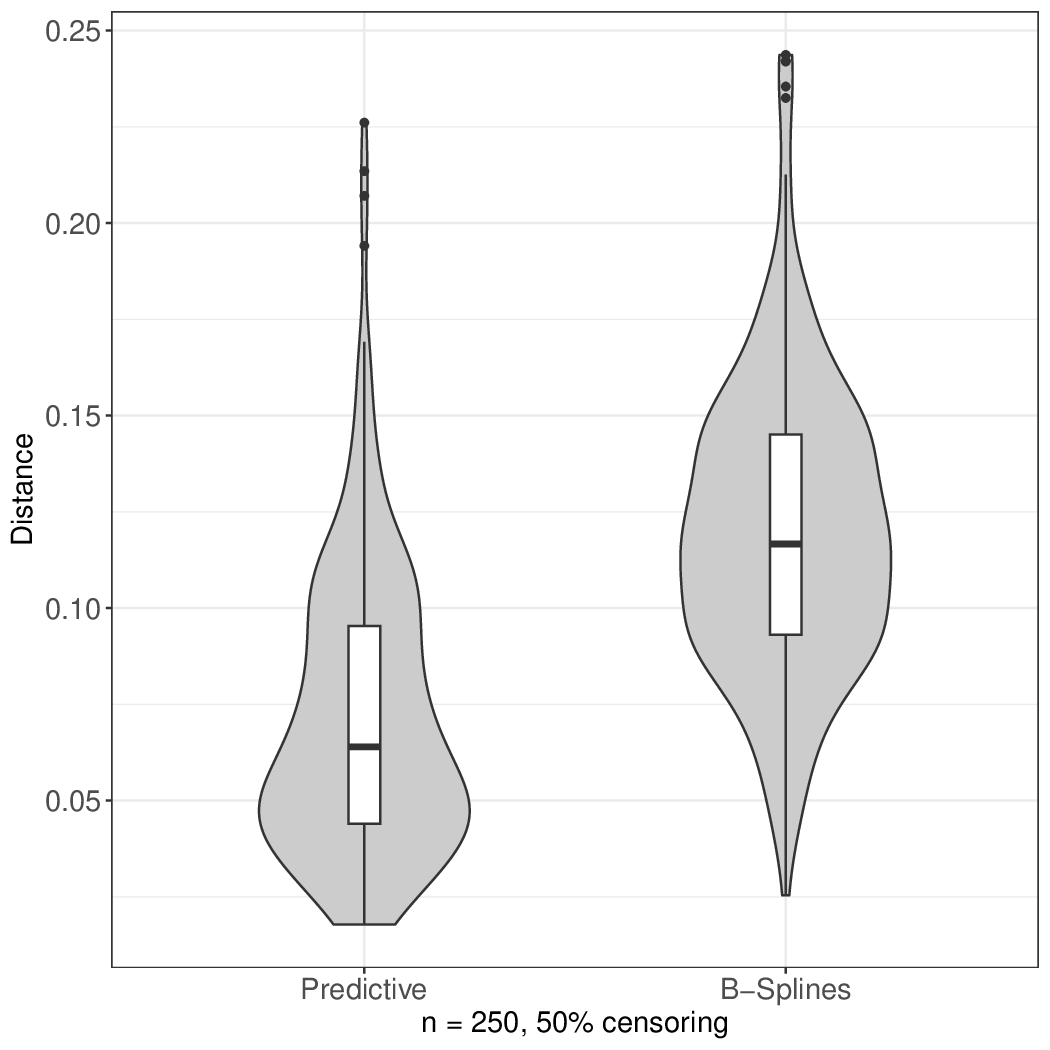} &
    \includegraphics[width = 6cm, height = 4cm]{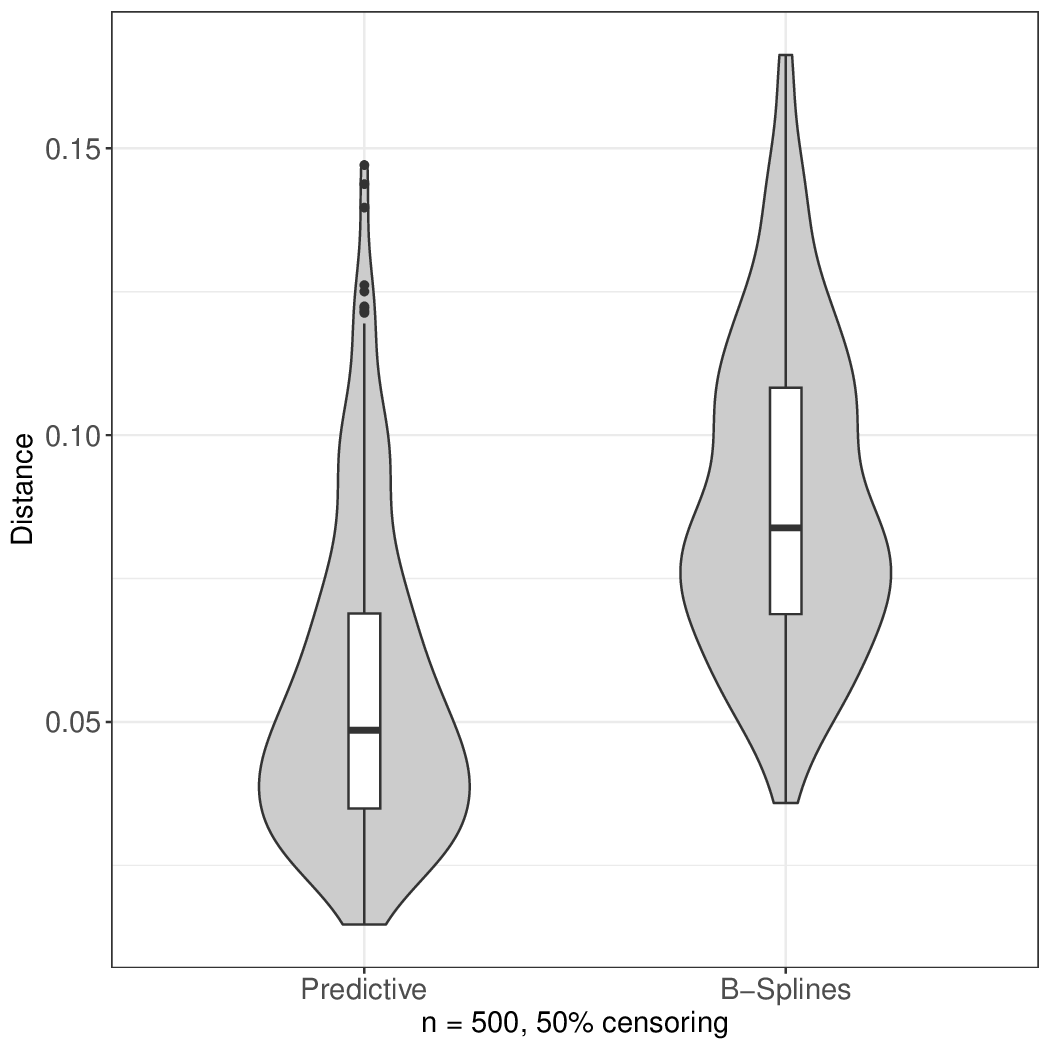} \\
    (a) $n=250$ & (b) $n=500$ \\
    \includegraphics[width = 6cm, height = 4cm]{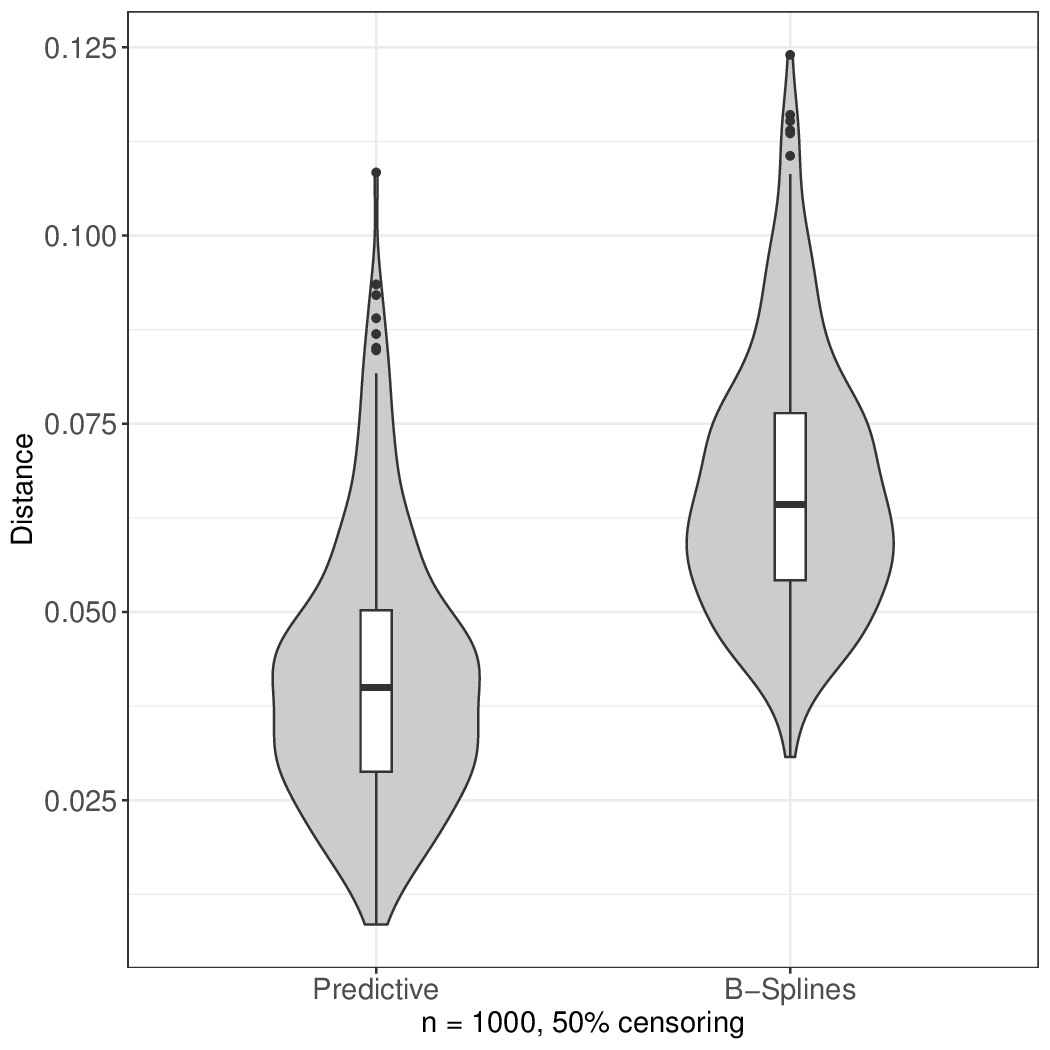} &
     \includegraphics[width = 6cm, height = 4cm]{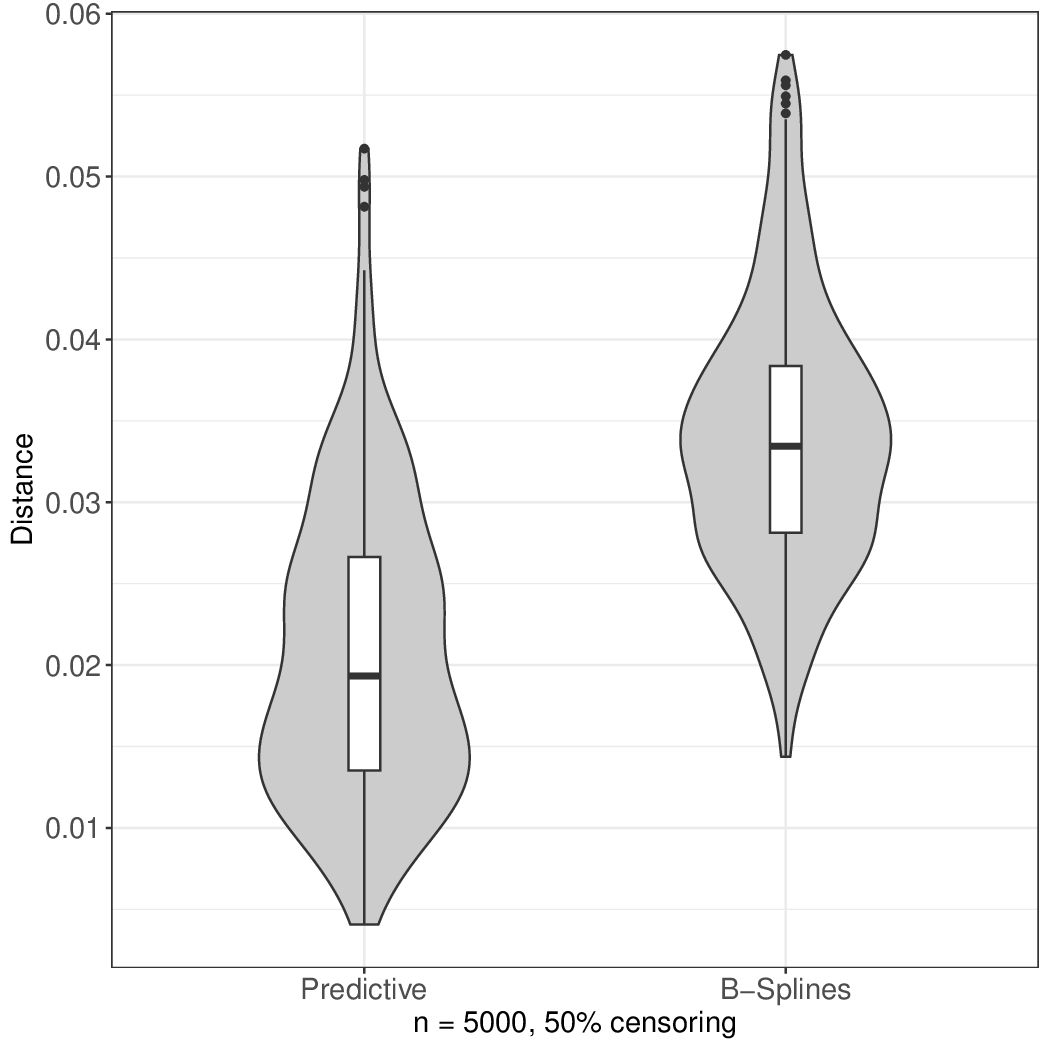}    \\
    (c) $n=1000$ & (d) $n=5000$
    \end{tabular}
    \caption{Simulation scenario 2, $50\%$ censoring: Distance between the posterior predictive hazards and the B-Spline estimator to the true generating model.}
    \label{fig:dist50}
\end{figure}

\clearpage

\bibliographystyle{SageH}
\bibliography{references}

% ...

\end{document}